\documentclass[11pt,a4paper]{article}
\usepackage{jcappub}
\usepackage{bm}
\usepackage{bbm}
\usepackage{amsmath}
\usepackage{etoolbox}
\usepackage{float}
\usepackage{mhchem}
\usepackage{gensymb}
\usepackage{color}
\usepackage{tikz}

\def\bea{\begin{eqnarray}}
\def\eea{\end{eqnarray}}
\def\be{\begin{equation}}
\def\ee{\end{equation}}

\definecolor{seagreen}{rgb}{0.18, 0.55, 0.34}

\usepackage{etoolbox}

\makeatletter
\gdef\@fpheader{}
\makeatother

\begin{document}

\makeatletter

\title{On the potential of pseudo-scalar dark energy}

\author[a,b]{Andrea Minotti,}
\author[a,b]{Yunzhi Wu,}
\author[a,b]{and Marco Regis}

\affiliation[a]{Dipartimento di Fisica, Universit\`{a} di Torino, via P. Giuria 1, I--10125 Torino, Italy}
\affiliation[b]{Istituto Nazionale di Fisica Nucleare, Sezione di Torino, via P. Giuria 1, I--10125 Torino, Italy}

\emailAdd{andrea.minotti@unito.it}
\emailAdd{yunzhi.wu@unito.it}
\emailAdd{marco.regis@unito.it}

\abstract{A cosmological pseudo-scalar field provides a compelling realization of dynamical dark energy (DE). If its coupling to photons is non-negligible, the cosmic microwave background acquires a rotation of its polarization plane, known as cosmic birefringence (CB). We present an extended analysis of several pseudo-scalar DE models and derive constraints on the parameters of their potentials by combining observations of the background expansion history with measurements of CB. We find that the axion-like potential constitutes a viable model only for large values of the anomaly coefficient. 
Scenarios in which the pseudo-scalar field rolls down a potential with quadratic, linear, or Ratra-Peebles forms can successfully explain DE and CB, with a symmetry-breaking scale close to the GUT scale.}

\maketitle

%\clearpage

\section{Introduction} \label{Sec:Introduction}
Recent results from cosmological surveys suggest that we may be approaching a cornerstone in modern cosmology. Several observational probes exhibit potential deviations from the concordance cosmological model~\cite{CosmoVerseNetwork:2025alb}. At present, however, these indications should be regarded as hints, as their statistical significance appears insufficient to robustly establish a departure from the vanilla $\Lambda$CDM paradigm.

Nevertheless, it is of paramount importance to carefully investigate the theoretical directions suggested by these emerging hints, in order to design and perform targeted tests and observations capable of either digging out genuine signals or definitively ruling out these possible deviations.

In this work, we focus on the possibility of dynamical dark energy, as recently suggested by the DESI Collaboration~\cite{DESI:2024mwx,DESI:2025zgx}. As well known, a scalar field can play the role of DE with an evolving equation of state~\cite{PhysRevD.37.3406,Zlatev:1998tr}, mimicking a cosmological constant when $\omega\simeq -1$. DESI measurements at very low redshift, combined with SNIa and CMB observations, appear to favor $\omega> -1$~\cite{DESI:2025zgx,DES:2025sig}, indicating a deviation from a pure cosmological constant and hinting at dynamical evolution.\footnote{At higher redshift, $\omega<-1$ might be mildly preferred~\cite{DESI:2025zgx}, although this conclusion is more analysis dependent~\cite{Cortes:2024lgw} and statistically less significant. Restricting ourselves to canonical scalar fields, we do not explore the $\omega<-1$ regime in this work.}

In particular, rather than considering pure scalar fields, we focus on pseudo-scalar fields, often referred to as axion-like particles (ALPs).
Due to their pseudo-scalar nature and their coupling to electromagnetism, ALPs induce different phase velocities for left- and right-circularly polarized photons, giving rise to a phenomenon known as cosmic birefringence (CB)~\cite{Carroll:1989vb,Harari:1992ea,Lue:1998mq}. This causes a rotation of the polarization plane, which can accumulate over cosmological distances in the ALP DE case, and become observable~\cite{Fujita:2020ecn}. 

Intriguingly, analyses of the polarization data from the CMB experiments Planck~\cite{Diego-Palazuelos:2022dsq,Eskilt:2022cff,Sullivan:2025wnv,Ballardini:2025apf,Remazeilles:2025wzd} and Atacama
Cosmology Telescope (ACT)~\cite{ACT:2025fju,Diego-Palazuelos:2025dmh} provided estimates of the CB angle away from zero. When the two datasets are combined, the statistical significance reaches $\gtrsim 4\sigma$~\cite{ACT:2025fju,Diego-Palazuelos:2025dmh,Yin:2025fmj}. In our statistical analysis, we will use 
$\beta_{\rm data} = 0.215 \pm 0.074$ deg from Ref.~\cite{Diego-Palazuelos:2025dmh}.

The work presented in this paper investigates different models of pseudo-scalar DE and constrains them using a combination of data probing the background dynamics of the Universe and measurements of CB.

Recently, the connection between axion DE and DESI results has been extensively investigated~\cite{Berbig:2024aee,Urena-Lopez:2025rad}.
On the other hand, only a few analyses included also CB. Ref.~\cite{Tada:2024znt} discussed DESI results and CB in the context of a specific realization of an axion-like thawing model. Ref.~\cite{Lee:2025yvn} focused on the hilltop axion scenario with domain walls and on a few benchmark models for the comparison with measurements. Ref.~\cite{Nakagawa:2025ejs} considered ALPs with a standard sine-Gordon potential, with no initial velocity, and compared cosmological predictions with DESI data through the effective Chevallier–Polarski–Linder (CPL) parameterization. 
Ref.~\cite{Lin:2025gne} also considered a standard axion-like potential, and employed a set of compressed information for the statistical analysis.
Ref.~\cite{Barman:2025ryg} computed the DE equation of state and CMB birefringence angle for a few benchmark cases of pseudo-scalar DE interacting with dark matter, and Ref.~\cite{Yin:2026gss} investigated early DE models.

Our work aims at providing a comprehensive overview of different pseudo-scalar DE models and at performing a statistical analysis employing the full cosmological likelihoods and detailed theoretical predictions.

The paper is organized as follows.
Sec.~\ref{Sec:Models} describes ALP DE and how cosmological predictions are computed making use of (a modified version of) the CLASS code~\cite{2011JCAP...07..034B}.
The data we employ are presented in Sec.~\ref{Sec:DM}, where the statistical analysis is also detailed.
Implications for the parameter space of different ALP DE models are discussed in Sec.~\ref{Sec:Results}, with conclusions drawn in Sec.~\ref{Sec:Conclusion}.
Throughout the paper, we will describe the Universe as flat, homogeneous, and isotropic.

\section{Cosmology of pseudo-scalar dark energy} \label{Sec:Models}
We introduce a pseudoscalar field $\phi$, which plays the role of DE, and it is described by a Lagrangian including a canonical kinetic term, a self-interaction potential $V$ and a coupling to electromagnetism:
\be
\mathcal{L}=\frac{1}{2}\partial^\mu \phi \partial_\mu \phi -V(\phi,\mu,f)-\frac{\alpha}{8\pi\,f}c_{\phi\gamma}\,\phi\,F^{\mu \nu}\tilde F_{\mu \nu}\;,
\label{eq:lagr}
\ee
where $f$ is the energy scale at which the effective description of the above Lagrangian breaks down, $\mu$ is an energy scale imprinted in the potential (e.g., given by the mass of the field in the case of axion-like and quadratic potentials), $\alpha$ is the fine-structure constant,  and $c_{\phi\gamma}$ is a dimensionless, model-dependent coefficient, parametrizing the anomaly, with typical values $\mathcal{O}(1)$.
The electromagnetic coupling of the field $\phi$ with $F^{\mu \nu}\tilde F_{\mu \nu}\propto \vec E \cdot \vec B$ is responsible for CB. A more compact way, often used, to write the coupling is through $g_{\phi\gamma}=\frac{\alpha}{2\pi\,f}c_{\phi\gamma}$.

In Sec.~\ref{Sec:Results}, we consider different theoretically well-motivated models of the potential $V$. Unless specified, we set $c_{\phi\gamma}=1$.
In our analysis, there are two free parameters associated with the Lagrangian of the pseudo-scalar DE, i.e., $f$ and $\mu$. Then, there are two initial conditions, one for the initial displacement of the field, $\phi_{\mathrm{ini}}$, and one for the velocity, $ \phi_{\mathrm{kick}}^\prime$, the latter describing a possible kick at a certain time of the evolution.

\subsection{CLASS implementation} \label{Ssec:CLASS}
For a given potential $V(\phi)$, the evolution of the background scalar field $\phi$ describing dark energy is computed (in a modified version of the CLASS code) by solving the field equation of motion: 
\begin{equation}
   \phi'' +2\mathcal{H}\phi'+a^2\frac{dV}{d\phi}=0
   \label{eq:eom}
\end{equation}
(with the prime denoting the derivative with respect to conformal time $\eta$ and $\mathcal{H} = a'/a$ with $a$ being the scale factor) together with the Friedmann equation:
\begin{equation}
   \mathcal{H}^2 = a^2H^2= a^2 \frac{8\pi G}{3}\left[ \rho_{\rm m}+\rho_{\rm r} +\rho_{\phi}\right]
\end{equation}
where:
\begin{equation}
    \rho_{\rm m}= \rho_{\rm b} +\rho_{\rm cdm} + \rho_{\rm ncdm}\ \ \ , \ \ \ 
   \rho_{\rm r} =  \rho_{\rm g} + \rho_{\rm ur}\;,
\end{equation}
with $\rm b=$ baryons, $\rm cdm=$ cold dark matter, $\rm ncdm=$ non-cold dark matter (i.e., one massive neutrino), $\rm g=$ photons and $\rm ur=$ ultra-relativistic species {(i.e., two massless neutrinos)}.
Individual energy densities for matter and radiation can be computed from the input parameters $\omega_{\rm b}, \  \omega_{\rm cdm},\ h, \ T_{\rm CMB}, \ N_{\rm ncdm}$ and $N_{\rm ur}$, where $\omega_b=\Omega_bh^2$ and $\omega_{\rm cdm}=\Omega_{\rm cdm}h^2$ are the physical baryon and cold dark matter densities, $h=H_0/(100\,{\rm km/s/Mpc})$ is the reduced Hubble constant, $T_{\rm CMB}$ is the current temperature of the CMB and $N_{\rm ncdm}$ and $N_{\rm ur}$ are the number of non-cold dark matter species and the number of ultra-relativistic species, respectively.
The background energy density of the scalar field, its pressure, and its equation of state are defined as: 
\begin{equation}
   \rho_{\phi}=\frac{\phi'^2}{2a^2} + V(\phi) \ \ \ , \ \ \ 
   P_{\phi} = \frac{\phi'^2}{2a^2} - V(\phi) \ \ \ , \ \ \ w_{\phi}=\frac{P_{\phi}}{\rho_{\phi}}.
\end{equation}
The evolution of the perturbations $\delta\phi=\phi-\bar{\phi}$ is computed by solving the perturbed Klein-Gordon equation in Fourier space:
\begin{equation}
   \delta\phi'' + 2\mathcal{H}\delta\phi' + \left( k^2 + a^2\frac{d^2V}{d\phi^2} \right)\delta\phi=-\frac{1}{2}h'\bar{\phi}',
\end{equation}
where again with the prime we denote the derivative with respect to conformal time and with $h'$ here we mean the derivative of the trace of the spatial metric perturbation in the synchronous gauge. In the synchronous gauge, in fact, $ds^2=a^2\left[-d\eta^2 + (\delta_{ij}+h_{ij})dx^idx^j\right]$.
\subsection{Cosmological predictions} \label{Ssec: Cosmological predictions}
From the evolution of the pseudo-scalar field we can immediately compute the predicted birefringence angle for CMB photons~\cite{Carroll:1989vb}:
\begin{equation}
   \beta = \frac{g_{\phi\gamma}}{2}\left(\phi_0-\phi_*\right) = \left(\frac{\alpha\,c_{\phi\gamma}}{4\pi f}\right)\left(\phi_0-\phi_*\right)\simeq 0.03^\circ\,c_{\phi\gamma}\frac{\Delta\phi}{f},
   \label{eq:angleCB}
\end{equation}
where $\phi_0$ and $\phi_*$ are the values of the field now and at recombination, respectively, and $\Delta\phi\equiv\phi_0-\phi_*$. Note that with $c_{\phi\gamma}\,\Delta\phi/f$ of a factor of a few, $\beta$ matches the observational estimates of CB from CMB experiments~\cite{Diego-Palazuelos:2022dsq,Diego-Palazuelos:2025dmh}, as mentioned in the Introduction, which is an interesting aspect of the ALP solution.
\\
Again making use of the CLASS code, we then compute different definitions of the cosmological distance:
\bea
   D_C(z) &\equiv& \int^{z}_{0}{\frac{c\,dz'}{H(z')}}\;,\quad\quad\quad
   D_L(z) \equiv (1+z) D_M(z)\;, \quad\quad\quad
   D_A(z) \equiv \frac{D_M(z)}{1+z}\;,\nonumber\\
   D_H(z) &\equiv& \frac{c}{H(z)}\;,\quad\quad\quad
   D_V(z) \equiv (z D_M(z)^2 D_H(z))^{1/3}\;, \quad\quad\quad D_M(z)=D_C(z).\nonumber
\eea
that are, respectively, the radial comoving distance, luminosity distance, angular diameter distance, Hubble distance, volume averaged distance, and transverse comoving distance in a flat, homogeneous and isotropic Universe.
They are affected by the evolution of $\phi$, and will be compared with the distance estimates obtained from the SNIa and BAO measurements.
\\
With our modified version of the CLASS code, we also compute the angular power spectra of cosmological density fluctuations, in the linear regime. We first derive the source functions $S(k,\tau)$ of perturbations by solving the linearized Boltzmann equations.
%\begin{equation}
%    S_T(k,\tau)\equiv g(\Delta_0+\psi)+(gk^{-2}\theta_b)' + e^{-k}(\phi'+\psi')+\text{polarization}.
%\end{equation}
Then the transfer functions $\Delta_{\ell}(k)$ are easily computed via: 
\begin{equation}
   \Delta_{\ell}(k)=\int_{\tau_{\rm ini}}^{\tau_0}S(k,\tau)j_{\ell}\left[ k(\tau_0-\tau)\right]d\tau.
\end{equation}
where $j_{\ell}$ denotes the spherical Bessel functions, $\tau_{0}$ is the conformal time today and $\tau_{\rm ini}$ is an initial conformal time where all relevant scales were sub-horizon ($k\tau_{\rm ini}<<1$).
The angular power spectra are then obtained by integrating the transfer functions, weighted by the primordial power spectrum $\mathcal{P}(k)$:
\begin{equation}
   C_{\ell}^{XY}=4\pi \int_{0}^{\infty}\Delta_{\ell,X}(k)\Delta_{\ell,Y}(k)\mathcal{P}(k)\frac{dk}{k}.
\end{equation}
The calculated $C_{\ell}$ power spectra will be compared with the Planck data~\cite{Planck2018,Tristram_2024}.

\section{Data and Methods} \label{Sec:DM}
In this Section, we first describe the datasets considered in our analysis, and then illustrate the statistical method used to compare models with data. 

\subsection{Datasets} \label{Sec:Data}

{\bf CMB power spectrum:}
In this work, we consider models of late-time dynamical DE, thus cosmological predictions deviate from the $\Lambda \mathrm{CDM}$ model only in the late Universe (far after recombination).
The main effect on the CMB is to modify the distance to last scattering, shifting the peaks of the CMB power spectra.
Moreover, the change in the equation of state influences the Integrated Sachs--Wolfe effect. 
These effects are accounted for through the CLASS computation.
In addition, the deviation in structure growth due to the modified expansion rate and equation of state can impact the CMB power spectra through gravitational lensing. This effect, which would require dedicated numerical cosmological simulations, is nevertheless expected to be very minor and will be disregarded (see also the discussion below).

The analysis is performed using the latest native (bundled) python NPIPE (PR4) CamSpec high-$\ell$ Temperature-Temperature (TT), E-mode Polarization (EE) and Temperature-E-mode (TE) power spectra
 likelihoods, which cover the multipoles from $\ell=30$ to $\ell=2500$~\cite{Tristram_2024,Torrado:2020dgo} for a total of 9915 data-points.
 For $\ell=2$ to $\ell=29$ we apply the Cobaya-native python implementation from Planck 2018 low-$\ell$ TT and EE likelihoods~\cite{Torrado:2020dgo} (56 data-points). Details of Planck CMB data can be found in Ref.~\cite{Planck2018}.

{\bf Supernova:}
The luminosity distance $D_L(z)$ of type Ia supernovae (SNIa) can be inferred from their light curves, through $m_B=M_B+25+5\log_{10}(D_L/{\rm Mpc})$, where $m_B$ and $M_B$ are the apparent and absolute magnitude, respectively. SNIa thus offer a probe of the expansion history of the Universe at low redshift, where DE has a major influence.
We employ the Pantheon$^+$ dataset, including 1701 light curves of 1550 distinct SNe Ia ranging from $z=0.001$ to $z=2.26$~\cite{Brout_2022}, and again using the Cobaya implementation of the likelihood~\cite{Torrado:2020dgo}.

{\bf Baryon Acoustic Oscillations:}
We use Baryon Acoustic Oscillations (BAO) data from the Dark Energy Spectroscopic Instrument (DESI) DR2~\cite{Abdul_Karim_2025}, obtained combining galaxies, quasars, and the Lyman-$\alpha$ forest as matter tracers. In our analysis, we compare the observed angles of BAO, $D_V/r_d, \, D_M/r_d$ and $D_H/r_d$, with $r_d$ being the comoving size of the sound horizon at the baryon drag epoch, at different redshifts in the range $z=[0.295,2.33]$ (for a total of 13 data-points), with the predictions from ALP DE models. 
Constraints on distances translate into constraint on the evolution of the Universe and thus on the equation of state $w(z)$.
BAO reside at linear clustering scales, thus they are free from systematic errors of the non-linear regime~\cite{Adame_2025}.

{\bf CMB Birefringence:}
We take the CB angle from ACT DR6~\cite{Diego-Palazuelos:2025dmh}:
$\beta_{\rm data} = 0.215 \pm 0.074$ deg as the fiducial value, and define the associated log likelihood as:
\begin{equation}
    \log\mathcal{L}_\beta=-\frac{(\beta_{\rm model}-\beta_{\rm data})^2}{2\sigma^2}
    \label{eq:likebeta}
\end{equation}

The global likelihood, combining the four probes, is obtained by multiplying the individual likelihoods described above, i.e., $\mathcal{L}=\mathcal{L}_{\rm CMB}\times\mathcal{L}_{\rm SN}\times \mathcal{L}_{\rm BAO}\times\mathcal{L}_\beta$.

\subsection{Statistical analysis} \label{Sec:Methods}
The parameter space is sampled using the Monte Carlo Markov Chain (MCMC) Metropolis sampler~\cite{Lewis:2002ah,Lewis:2013hha}. The results of the sampling are analysed with GetDist, and both tasks are performed through the Cobaya code~\cite{Torrado:2020dgo}.
For the statistical analysis, we adopt the likelihood depicted in the previous Section combined with the priors reported in Table~\ref{table:priors}.
When presenting results, we will show the 1D and 2D posterior distributions, highlighting 68\% and 95\% credible regions. To test the robustness of our statistical analysis, we also report the same contours but using a frequentist approach, i.e., we show the 1D and 2D profile likelihood distributions.

As discussed in Sec.~\ref{Sec:Models}, theoretical predictions are computed running the CLASS code~\cite{2011JCAP...07..034B} as Boltzmann equations-solver.
We do not include effects on the non-linear clustering from dynamical DE. To estimate the impact of this assumption, we run CLASS for some examples of the models considered below with and without the inclusion of the $\Lambda$CDM non-linear clustering, calculated with Halofit~\cite{Smith:2002dz}. The biggest changes occur in the CMB lensing, which mainly depends on low-redshift non-linear structures, but with a small impact on the $C_\ell^{TT}$ and $C_\ell^{EE}$ considered in our analysis, and negligible changes in the inferred posterior distributions. We therefore believe that it is legitimate to assume that a possible correction to this term arising from dynamical DE (which would be much smaller than the complete removal of non-linear clustering tested above) can be disregarded in our statistical analysis.

With respect to the concordance flat-$\Lambda$CDM model, our scenario replaces a cosmological constant with a dynamical DE described by a pseudo-scalar field.
In the statistical analysis, the parameter describing the DE abundance $\Omega_{\Lambda}$, with $w=-1$ in $\Lambda$CDM, is thus replaced by a few parameters describing the dynamical DE potential. Their explicit definitions and the expression of the potential for the different models considered in this work will be introduced in Sec.~\ref{Sec:Results}. 
The parameters sampled in the MCMC and shared by both the $\Lambda$CDM and the ALP DE scenarios are:
\begin{equation*}
    A_s, n_s, \omega_b, \omega_{\mathrm{cdm}}, \tau_{\mathrm{reio}},
\end{equation*}
which represent, respectively, the primordial scalar amplitude, the scalar spectral index, the physical baryon and cold dark matter densities, and the reionization optical depth.
In addition to these, we sample the ALP DE parameters:
\begin{table}[H]
    \centering
    \begin{tabular}{c||c}
           Axion-like: & $f$, $\mu$, $\phi'_{ \mathrm{kick}}$, $\phi_{\mathrm{ini}}$, $c_{\phi\gamma}$\\
          Linear $+$: &$f$, $\mu$, $\phi'_{ \mathrm{kick}}$, $\phi_{\mathrm{ini}}$  \\
          Linear $-$: & $f$, $\mu$, $\phi_{\mathrm{ini}}$\\
          Quadratic: & $f$, $\mu$, $\phi_{\mathrm{ini}}$\\
          Ratra-Peebles: & $f$, $\mu$ , $\phi_{\mathrm{ini}}$
    \end{tabular}
    \caption{Parameters of the pseudo-scalar DE models sampled in MCMC for each potential.}
    \label{tab:placeholder}
\end{table}
The initial displacement is set in the very early Universe, and the dependence on the exact redshift chosen is negligible, since the field is frozen by Hubble dragging. Instead, we set a field velocity (a ``kick") well after recombination, otherwise it would have no effects, i.e., the velocity would be quickly washed out by the Hubble dragging.
For the sake of definiteness, we set $z_{\rm kick}=9$, and discuss other possibilities in the following.

\begin{table}[h!]
  \centering
  \begin{tabular}{|c|c|c|c|c|c|} 
  \hline
  \multicolumn{6}{|c|}{\textbf{Priors}}\\
\hline
\textbf{Parameters} & $f $ & $\mu $ & $\phi'_{ \mathrm{kick}}$ &  $\phi_{\mathrm{ini}}$ &  $c_{\phi\gamma}$ \\ 
\hline
    \textbf{Axion-like} & $[0,M_{\mathrm{P}}/\pi]$& $[0.1,10]\,H_0$ & $[0,80]\frac{M_{\mathrm{P}} H_0}{\sqrt{8\pi}}$ &$[-M_{\mathrm{P}},0]$  &$[1,100]$ \\ 
  \hline
  \textbf{Linear} & $[0,M_{\mathrm{P}}]$& $[0,1]\sqrt{\frac{M_{\mathrm{P}}}{{\rm Mpc}}}$ &$[0,100]\frac{M_{\mathrm{P}} H_0}{\sqrt{8\pi}}$  &$[0,M_{\mathrm{P}}]$  & 1 \\ 
  \hline
  \textbf{Minus Linear} & $[0,M_{\mathrm{P}}]$& $[0,1]\sqrt{\frac{M_{\mathrm{P}}}{{\rm Mpc}}}$ & 0 &$[-M_{\mathrm{P}},0]$  & 1 \\ 
  \hline
  \textbf{Quadratic} & $[0,M_{\mathrm{P}}]$& $[0.1,10]\,H_0$ & 0 &$[-M_{\mathrm{P}},0]$  & 1 \\ 
  \hline
   \textbf{Ratra-Peebles} & $[0,M_{\mathrm{P}}]$& $[0,0.017]\sqrt{\frac{M_{\mathrm{P}}}{{\rm Mpc}}}$ & 0 &$[0,M_{\mathrm{P}}]$  & 1 \\ 
  \hline
\end{tabular}
  \caption{Prior range for the parameters of the ALP DE models sampled in the MCMC. $M_{\mathrm{P}}$ is the Planck mass. }
\label{table:priors}
\end{table}

\section{Results} \label{Sec:Results}
In this Section we introduce different potentials for pseudo-scalar DE, and, for each of them, we perform the statistical analysis described above.

Let us remind the physical conditions our models try to meet: a) we need the field to roll-down at recent times to explain DE evolution; b) we need $c_{\phi\gamma}\,\Delta\phi>0$ in order to correctly reproduce the sign of the CB angle, see Eq.~\ref{eq:angleCB}.
In the following, we assume $c_{\phi\gamma}>0$, which implies a positive field excursion $\Delta\phi>0$ , i.e., the field has increased from recombination to the present day.
We will also comment about the scenarios with $c_{\phi\gamma}\,\Delta\phi<0$ leading to equivalent phenomenology.

\subsection{Axion-like potential}\label{sec: Axion-like potential}
We start considering an axion-like potential with the form:
\begin{equation}
    V(\phi)=m^2f^2\left[1-\cos\left(\frac{\phi}{f}\right)\right],
    \label{eq:cosV}
\end{equation}

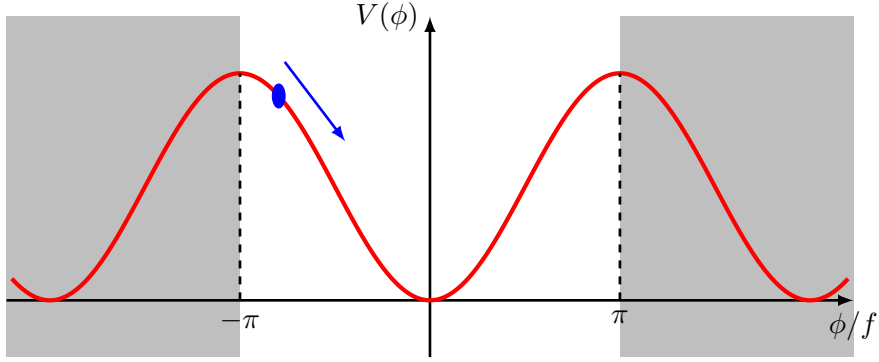
\begin{figure}[H]
    \centering
    \begin{tikzpicture}[xscale=0.8, yscale=1.5, >=latex]
    \fill[gray!50] (3.14, -0.5) rectangle (7, 2.5);
    \fill[gray!50] (-7, -0.5) rectangle (-pi, 2.5);
    \draw[->, line width=1pt] (-7,0) -- (7,0) node[below] {$\phi/f$};
    \draw[->, line width=1pt] (0,-0.5) -- (0,2.5) node[left] {$V(\phi)$};
    \draw[color=red, ultra thick, smooth, samples=200] 
        plot[domain=-2.2*pi:2.2*pi] (\x, {1-cos(\x r)});
    \node[below] at (-3.14,0) {$-\pi$};
    \node[below] at (3.14,0) {$\pi$};
    \draw[dashed, line width=1pt] (-3.14,0) -- (-3.14,2);
    \draw[dashed, line width=1pt] (3.14,0) -- (3.14,2);
    \filldraw[blue] (-2.5, 1.8) circle (3pt);
    \draw[->, line width=1pt, color=blue] (-2.4,2.1) -- (-1.4,1.4);
    \end{tikzpicture}
    \centering
    \caption{Axion-like potential.}
    \label{fig:ALPs_Potential}
\end{figure}
Here $\mu=m$ is the ALP mass, which is expected to be somewhat larger than $H_0\simeq 10^{-33}$ eV, i.e., not smaller than that so that the ALP field is not frozen and can lead to an evolution at late times, and not much larger so that it can play the role of DE and do not behave as a matter component. Note from Table~\ref{table:priors} that the prior for $f$ in this scenario has a factor $\pi$ of difference with respect to the other cases. This is to avoid crossing a maximum of the cosine, as depicted in Fig.~\ref{fig:ALPs_Potential}. 

Another condition that will be common to all the different potentials, is that the energy density of DE has to be of the order of meV$^4$, as well-known.
For the axion-like potential, this translates into a constraint for $m^2\,f^2$ which explains the degeneracy between $f$ and $m$, and, given the range of $m$ mentioned above, this leads the breaking scale $f$ to be of the order of a fraction of $M_{\mathrm{P}}$. 

With $c_{\phi\gamma}$ set to 1, we found that the CB likelihood and the cosmological likelihoods
(Planck CMB + DESI BAO + Pantheon$^+$) are not compatible. Essentially, once the parameters of the potential are set to fit the cosmological probes, the axion-photon coupling turns out to be too low to provide a CB angle matching its measurement. In other words, $f$ is large due to the above argument, and Eq.~\ref{eq:angleCB} would require a trans-planckian
evolution of $\phi$ to match the observed CB angle $\beta$.
In order to solve this issue, we ran a set of MCMC with $c_{\phi\gamma}\in [1,100]$. 
We consider two distinct scenarios: one where the field only rolls down the potential (``pure thawing") and a second one where the field first rolls up, then it stops, and subsequently rolls down (``freezing and thawing").

\subsubsection{Pure thawing scenario}\label{sec:rolldown}
\begin{figure}[H]
    \centering
\includegraphics[width=0.49\textwidth]{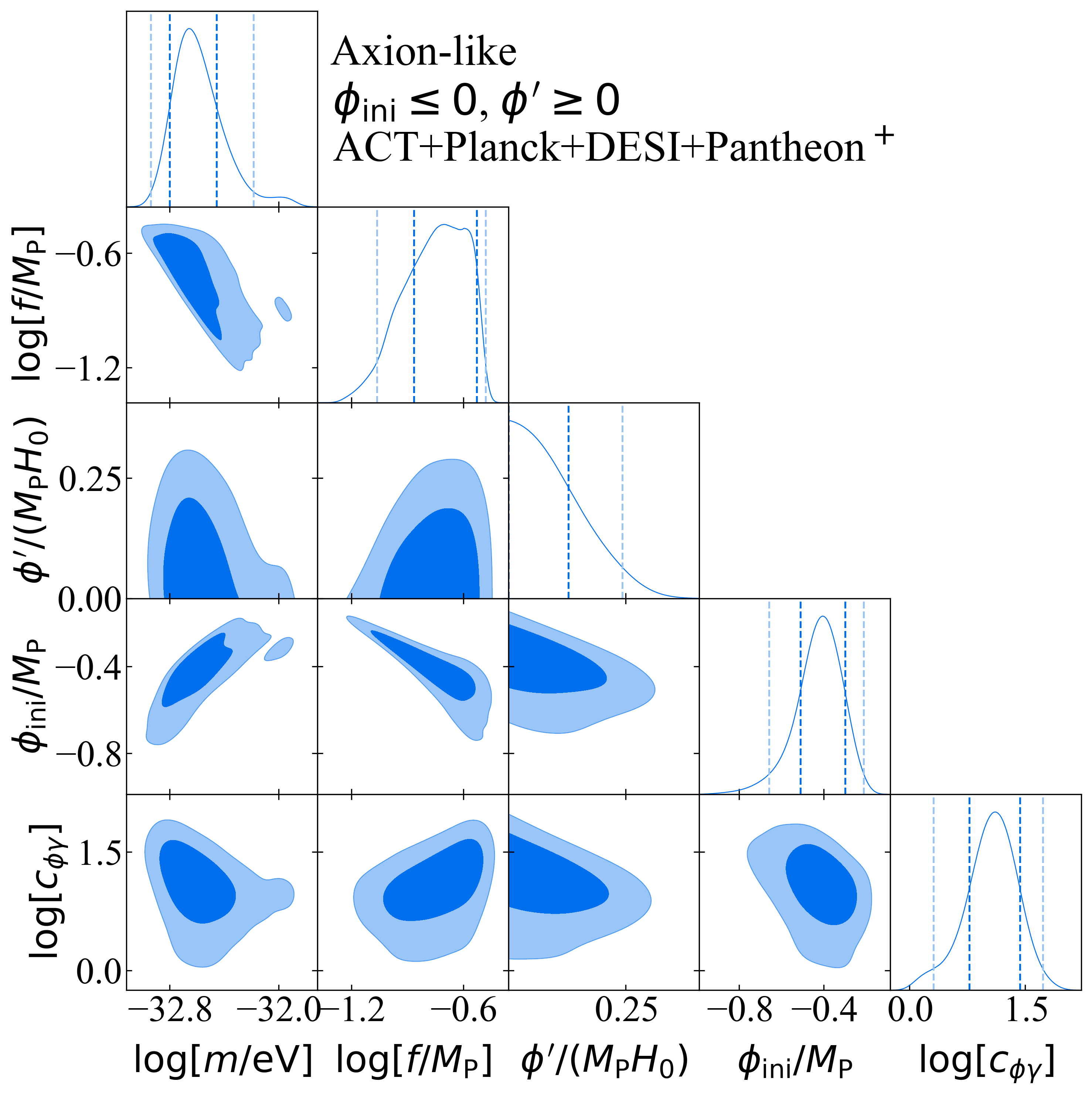}
\includegraphics[width=0.49\textwidth]{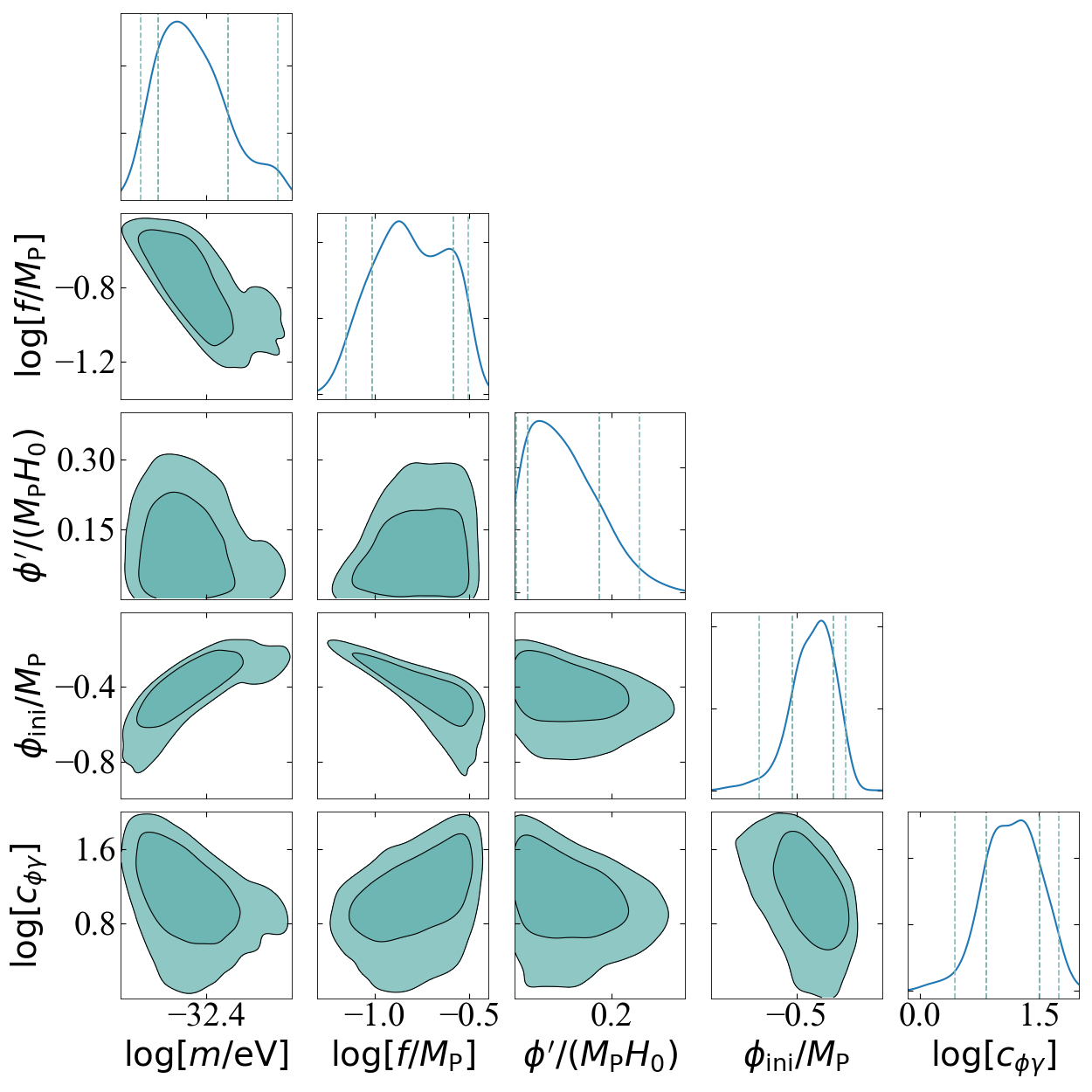}
    \centering
    \caption{Constraints on the parameters describing the pseudo-scalar DE model with axion-like potential in the pure thawing scenario. Left: In all panels, the 2D contours refer to the 68\% and 95\% credible regions of the 2D posterior distribution, while the first panel of each column provides the 1D posterior distribution of the individual parameter (with dashed lines referring to the 68\% and 95\% intervals). Right: Same as left, but for the confidence regions of the profile likelihood distribution. }
    \label{fig:AL}
\end{figure}

Here we consider a pure roll-down scenario. Fig.~\ref{fig:AL} shows the corner plot reporting the 1D and 2D posterior and profile likelihood distributions of the model parameters. 

The initial displacement $\phi_{\mathrm{ini}}$ has to be negative, so to provide $\Delta\phi>0$, i.e., the correct sign for the CB angle (for a positive $c_{\phi\gamma}$).
The scenario with $\phi_{\mathrm{kick}}^\prime>0$ is such that the field rolls down the potential in a faster way than without the velocity kick. This extra degree of freedom does not seem to help the model fit the data better (i.e., the data do not suggest $\omega\gg-1$). This is shown by Fig.~\ref{fig:AL} where $\phi_{\mathrm{kick}}^\prime$ turns out to be compatible with zero, and only an upper bound can be derived.
A case with $\phi_{\mathrm{kick}}^\prime<0$ has not been considered, as it is at odd with CB.

Note that there is a correlation between $\phi_{\mathrm{ini}}$ and $f$, such that $-\pi\lesssim\phi_{\mathrm{ini}}/f\lesssim -\pi/2$. This means that the preferred starting point is not far from the hilltop, or, at least, before the inflection point of the potential, in order to be on a slope reproducing the evolution required by the data.
Because of the aforementioned degeneracy between $m$ and $f$, the initial displacement shows in turn a degeneracy also with $m$. 

With the introduction of $c_{\phi\gamma}$, all the other parameters decouple from CB and are set by DE data alone. The peak of the posterior and profile likelihood distributions of $\log[c_{\phi\gamma}]$ are away from zero, and we find $c_{\phi\gamma}= 11.48^{+15.43}_{-5.46}$. 
Note that the field excursion (that can be derived from Eq.~\ref{eq:angleCB}) is relatively small, $\Delta\phi/\phi_{\mathrm{ini}}\approx0.25$.

All mean values and 68\% credible intervals of the model parameters, as well as the best-fit values and 68\% confidence intervals, are reported in Table~\ref{table:params}. It is remarkable to note that the results are not significantly dependent on the statistical approach, and the Bayesian and frequentist methods provide very compatible outcomes (this will be true also for the other models considered in the following).

Fig.~\ref{fig:AL} shows that the preferred ALP mass is larger than $H_0$, which leads to DE evolution at late times. The velocity kick drives a significant variation at high redshift, as we will discuss in Sec.~\ref{sec:comparison}. This is however not relevant for cosmology, since the ALP DE density is subdominant at the epoch of the kick. One can easily check that the constraints on $\omega(z)$ from the binning reconstruction approach on DESI data \cite{Abdul_Karim_2025} are very loose at $z\gtrsim 2$ and the integrated effect on the distance to last scattering surface of the CMB is clearly also very small.
Then due to the Hubble dragging, the kinetic energy decreases so that the parameter $\omega$ of the equation of state reaches values around $-1$. When the dragging becomes weaker than the evolution driven by the potential, the kinetic energy starts to increase again, and $\omega$ grows to about $-0.9$. This evolution makes ALP DE different from the case of the cosmological constant where $\omega=-1$ is constant, and our analysis points towards a statistical preference for such evolution, as we will discuss later on.

Flipping the sign of $\phi_{\mathrm{ini}}$, $\phi_{\mathrm{kick}}^\prime$, and $c_{\phi\gamma}$ would lead to the same conclusions discussed here, i.e., to exactly the same corner plots, just with different signs.

\subsubsection{Freezing and thawing scenario}\label{sec:kicked}
\begin{figure}
    \centering
\includegraphics[width=0.49\textwidth]{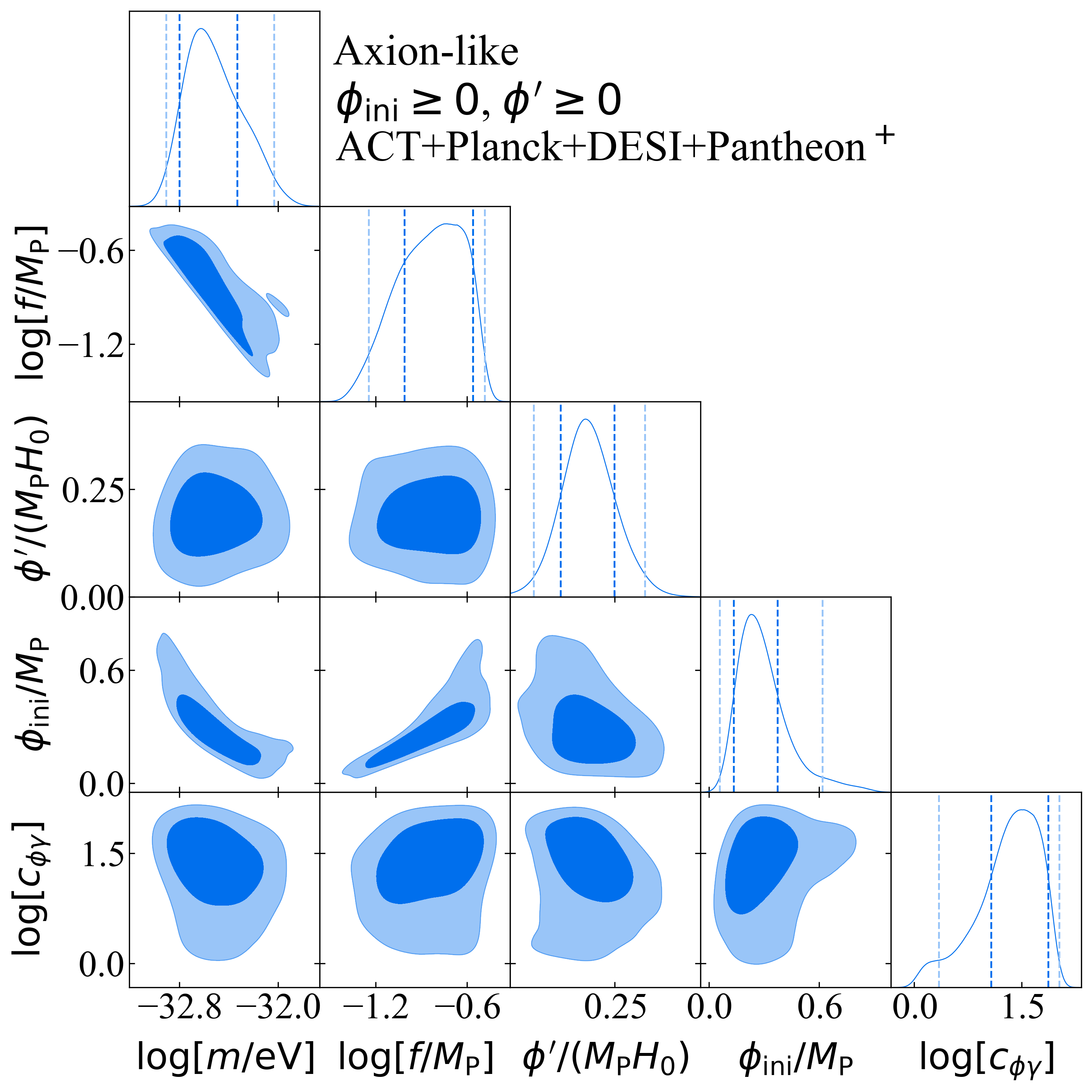}
\includegraphics[width=0.49\textwidth]{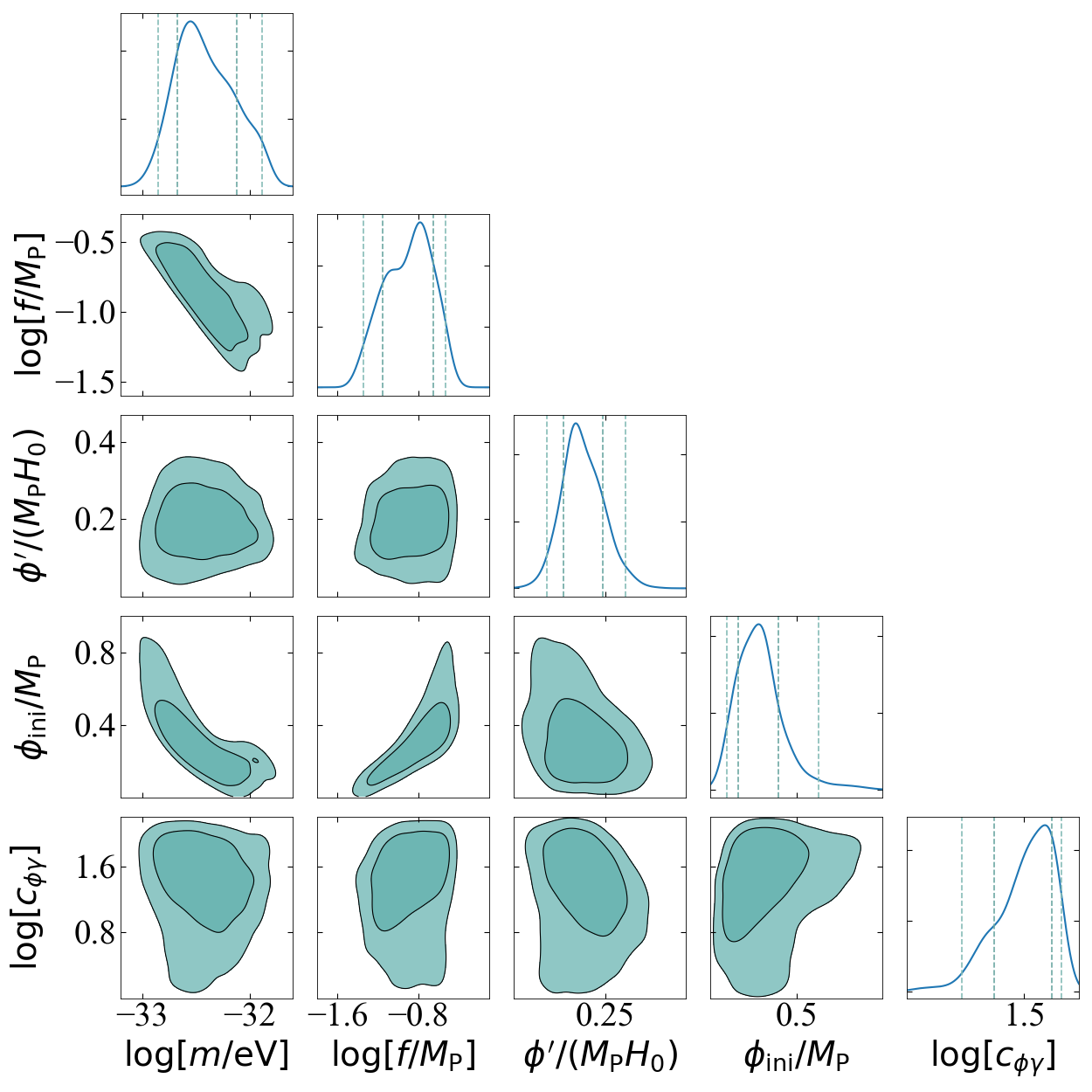}
    \centering
    \caption{Same as Fig.~\ref{fig:AL}, but for the freezing and thawing scenario.}
    \label{fig:kicked}
\end{figure}

A scenario with $\phi_{\mathrm{ini}}>0$ and $\phi_{\mathrm{kick}}^\prime>0$ is such that the field rolls up the potential, then reaches a maximum, and subsequently rolls down. 
Clearly, it has distinctive features with respect to what has been discussed above.
This model has also been called ``kicked quintessence"~\cite{Berbig:2024aee} and can meet the physical requirements outlined in the introduction of Sec.~\ref{Sec:Results}, if the value of the field today is greater than $\phi_{\mathrm{ini}}$ (again, assuming $c_{\phi\gamma}>0$).
From Fig.~\ref{fig:kicked}, we see that this scenario is a viable possibility, but at the price of a quite large anomaly coefficient, in order to fit CB. We find $c_{\phi\gamma}=19.95^{+58.93}_{-8.42}$.
As in the previous case, the details of the initial kick are not relevant for the cosmological evolution. Indeed, the field has to stop its evolution, recovering the limit $\omega=-1$ at $z\gtrsim2$, until the potential makes it rolling down at late-times, leading to the evolution suggested by the DESI data.

A simultaneous flip of the sign of $\phi_{\mathrm{ini}}$, $\phi_{\mathrm{kick}}^\prime$, and $c_{\phi\gamma}$ would lead to the same conclusions.
Since this scenario does not improve the picture with respect to the pure thawing case, for brevity, we report only the latter in Table~\ref{table:params} and Sec.~\ref{sec:comparison}.

\subsection{Linear potential}\label{sec: Linear potential}
\begin{figure}
    \includegraphics[width=0.49\textwidth]{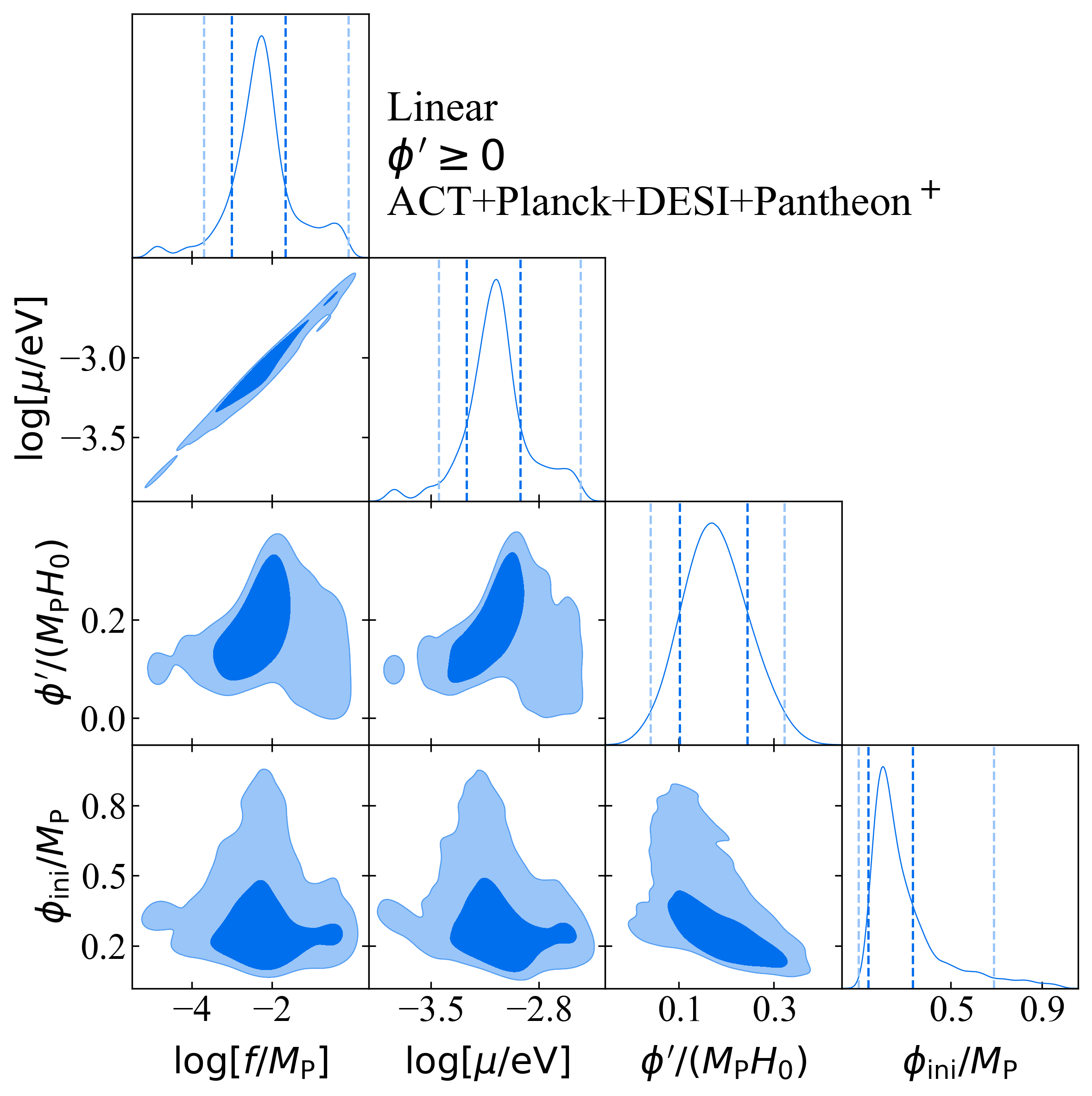}
 \includegraphics[width=0.49\textwidth]{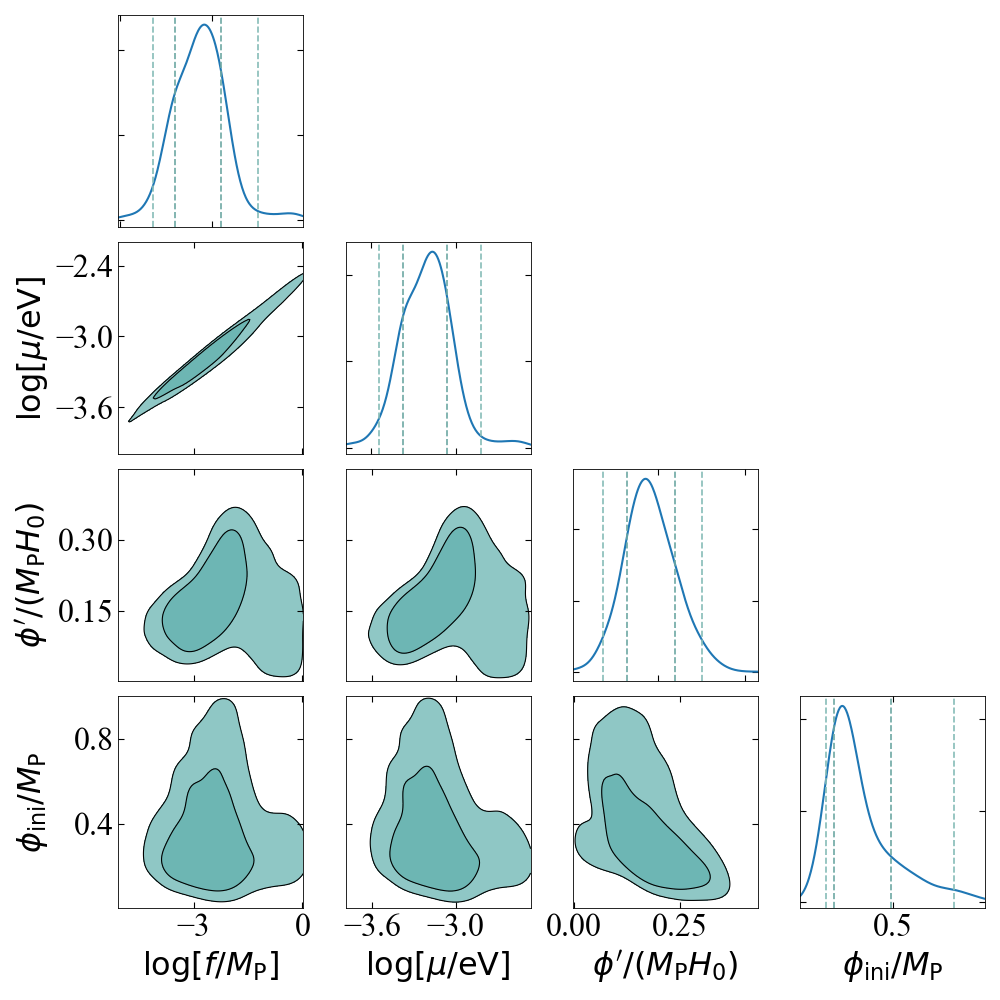}
   \caption{Same as Fig.~\ref{fig:AL}, but for the case of the linear potential with positive slope.}
    \label{fig:LIp}
\end{figure}

A linear potential for ALPs appears in some string theory scenarios \cite{Gasparotto:2022uqo, Panda:2010uq, Gasparotto:2023psh, Flauger:2009ab, Cicoli:2023opf}, and it can be parameterized as follows:
\begin{equation}
    V(\phi)=\mu^{4}\frac{\phi}{f},
    \label{eq:linpot}
\end{equation}
\begin{figure}[H]
    \centering
    \begin{tikzpicture}[xscale=0.6, yscale=1.2, >=latex]
     \fill[gray!50] (5, -2) rectangle (7, 2.5);
    \fill[gray!50] (-7, -2) rectangle (-5, 2.5);
    \draw[->, line width=1pt] (-7,0) -- (7,0) node[below] {$\phi$};
    \draw[->, line width=1pt] (0,-2) -- (0,2.5) node[left] {$V(\phi)$};
    \draw[color=red, ultra thick, smooth, samples=200] 
        plot[domain=-7:7] (\x, {0.2*\x});
    \node[below] at (-5,0.44) {$-M_{\mathrm{P}}$};
    \node[below] at (5,0) {$M_{\mathrm{P}}$};
    \draw[dashed, line width=1pt] (-5,-1) -- (-5,0);
    \draw[dashed, line width=1pt] (5,0) -- (5,1);
    \filldraw[blue] (2, 0.4) circle (3pt);
    \filldraw[green!80] (4, 0.8) circle (3pt);
    \draw[->, line width=1pt, color=blue] (2,0.6) -- (4,1.0);
    \draw[->, line width=1pt, color=green] (4,1.2) -- (3,1.0);
    \end{tikzpicture}
    \centering
    \caption{Linear potential with positive slope.}
    \label{fig:Linear_Potential}
\end{figure}
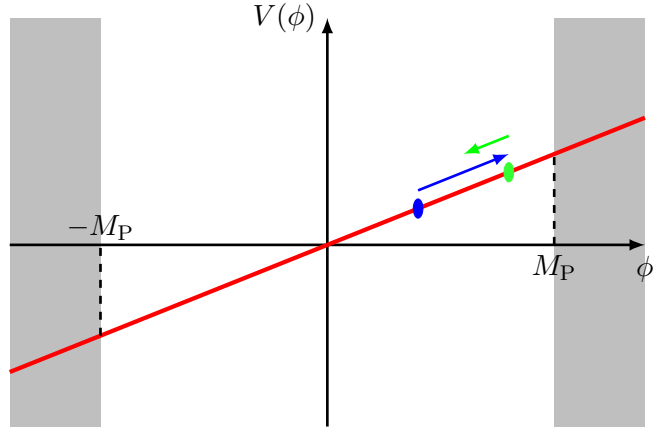
The only viable possibility with a roll-down at recent times and $\Delta\phi>0$ is with a positive initial velocity $\dot{\phi}_{in}$. In this way, the potential first increases and then the field rolls down but to a value larger than at early times, as illustrated in Fig.~\ref{fig:Linear_Potential}.
Thus, the explanation of CB is at the origin of having $\phi_{\mathrm{kick}}^\prime$ away from zero in Fig.~\ref{fig:LIp}, as for the case of kicked quintessence discussed in Sec.~\ref{sec:kicked}. The parameters $f$ and $\mu$ have a significant degeneracy due to the definition of potential in Eq.~\ref{eq:linpot}, and to the already mentioned requirement of having the DE density $\sim {\rm meV}^4$.
Since $f$ and $\phi_{\mathrm{ini}}$ are of the same order, $\mu$ comes out to be at the meV scale. Note that $f$ can be significantly smaller than in the ALP case and can match typical GUT energy scales.

Again, these conclusions are valid also for a model with simultaneous flip of the sign of $\phi_{\mathrm{ini}}$, $\phi_{\mathrm{kick}}^\prime$, and $c_{\phi\gamma}$. 

\subsubsection{Linear potential with negative slope}\label{sec: Minus Linear potential}
Let us consider again a linear potential but with the opposite slope (called ``minus linear" in figures and tables):
\begin{equation}
    V(\phi)=-\mu^{4}\frac{\phi}{f},
\end{equation}
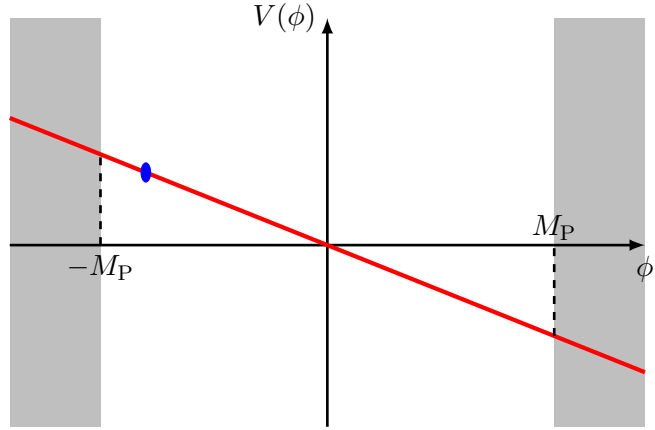
\begin{figure}
    \centering
    \begin{tikzpicture}[xscale=0.6, yscale=1.2, >=latex]
      \fill[gray!50] (5, -2) rectangle (7, 2.5);
    \fill[gray!50] (-7, -2) rectangle (-5, 2.5);
    \draw[->, line width=1pt] (-7,0) -- (7,0) node[below] {$\phi$};
    \draw[->, line width=1pt] (0,-2) -- (0,2.5) node[left] {$V(\phi)$};
    \draw[color=red, ultra thick, smooth, samples=200] 
        plot[domain=-7:7] (\x, {-0.2*\x});
    \node[below] at (-5,0) {$-M_{\mathrm{P}}$};
    \node[below] at (5,0.44) {$M_{\mathrm{P}}$};
    \draw[dashed, line width=1pt] (5,-1) -- (5,0);
    \draw[dashed, line width=1pt] (-5,0) -- (-5,1);
    \filldraw[blue] (-4, 0.8) circle (3pt);
    \end{tikzpicture}
    \centering
    \caption{Linear potential with negative slope.}
    \label{fig:Minus_inear_Potential}
\end{figure}

\begin{figure}
    \centering
    \includegraphics[width=0.5\linewidth]{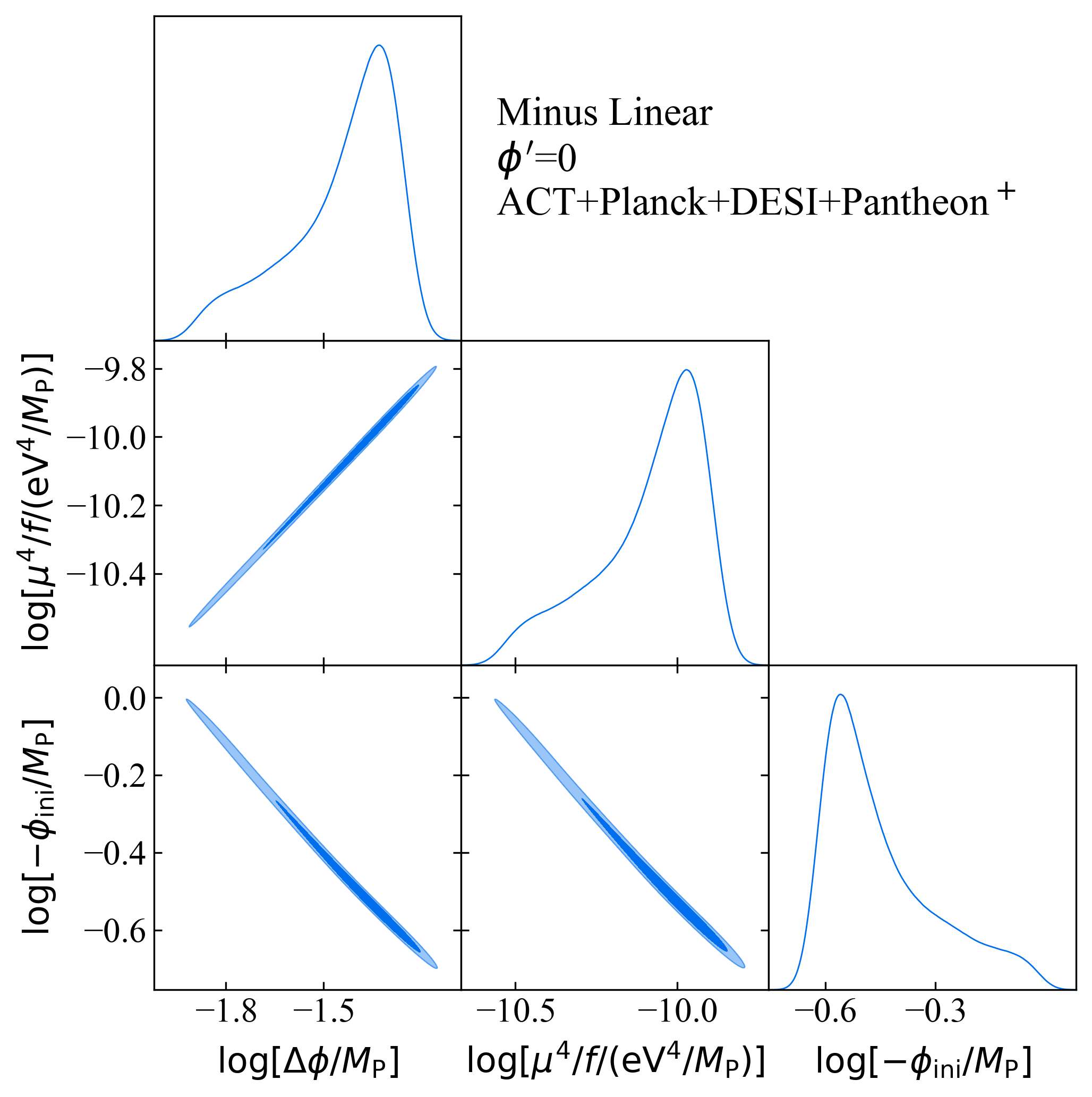}
    \caption{Corner plot of relevant quantities to understand degeneracies in the parameter space for the linear potential with negative slope. See text for more details.}
    \label{fig:invlin_mufphi}
\end{figure}

\begin{figure}
    \centering
\includegraphics[width=0.49\textwidth]{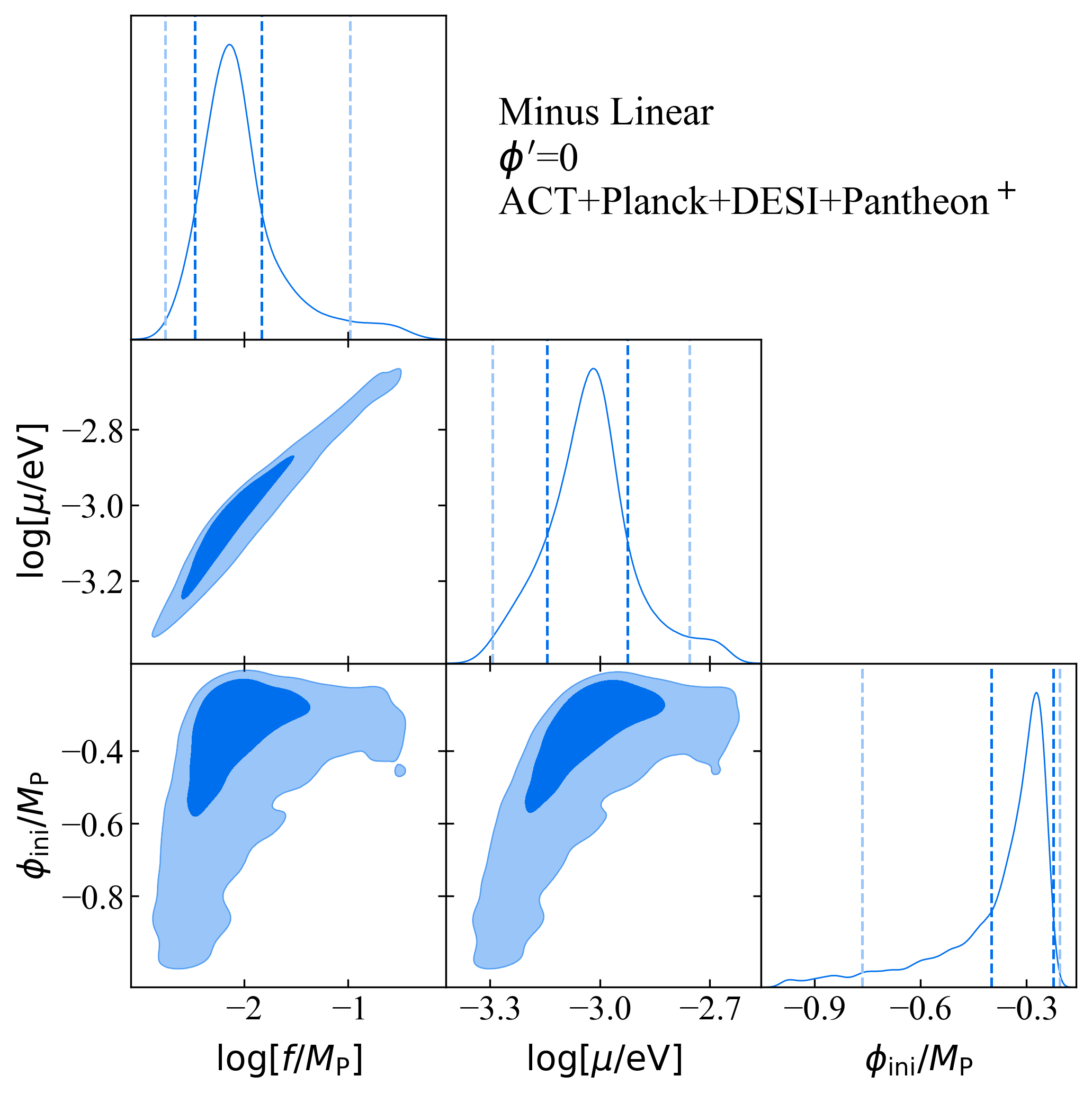}
\includegraphics[width=0.49\textwidth]{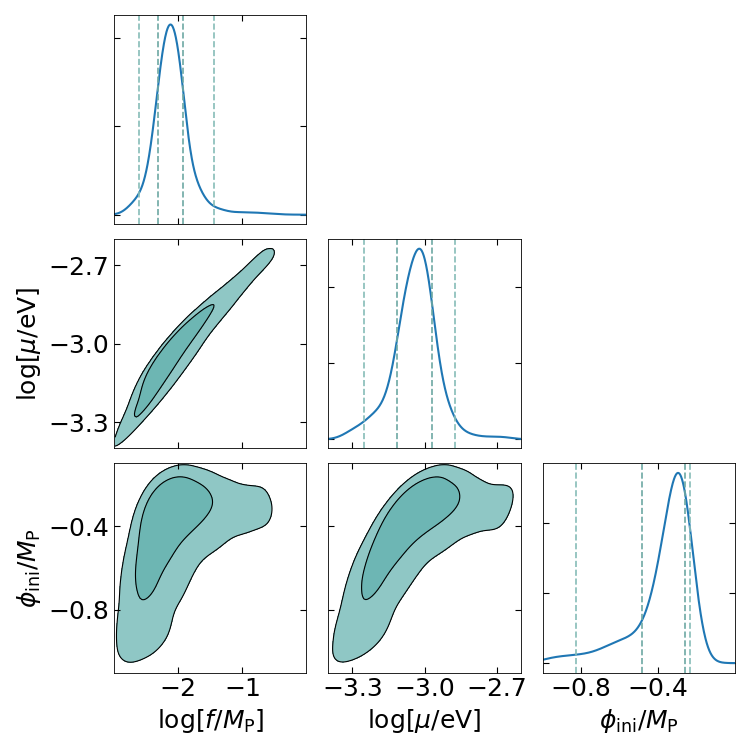}
    \centering
    \caption{Same as Fig.~\ref{fig:AL}, but for the case of the linear potential with negative slope.}
    \label{fig:LIm}
\end{figure}

In this case, there is no need of an initial velocity $\phi_{\mathrm{kick}}^\prime$ to explain the positive CB angle. On the other hand, a negative initial value $\phi_{\mathrm{ini}}<0$ is needed for the field energy density to be positive. 
As in the other scenario, the stronger constraint comes from the DE energy density, i.e., $\mu^4|\phi_{\mathrm{ini}}|/f\approx {\rm meV}^4$, which leads to a degeneracy between the slope $\mu^4/f$ and $\phi_{\mathrm{ini}}$. This is partially broken by the DE evolution and CB in the following way.
The equation of motion Eq.~\ref{eq:eom} has a simple solution if the dragging term can be ignored: $\Delta\phi(t)=(\mu^4/2f) \,\Delta t^2$.
Large $|\phi_{\mathrm{ini}}|$, i.e., low $\mu^4/f$ (for what said above), means low $\Delta\phi$ (the expansion rate is very constrained for all models, so $\Delta t$ does not change). Such degeneracy trends are reported in Fig.~\ref{fig:invlin_mufphi}. Low $\Delta\phi$ implies low $f$ in order to fit CB, see Eq.~\ref{eq:angleCB} and this explains the bottom-left panel of Fig.~\ref{fig:LIm}. 
The parameter $\mu$ does not enter the CB determination, and thus inherits the degeneracy with $\phi_{\mathrm{ini}}$ related to DE energy density, with the allowed range set by the requirement of DE evolution.
The preferred values of $\mu$ and $f$ are in a similar ball-park as in the model with positive slope.
For both potentials, the field excursion is small (which is also the reason why $f$ is relatively small). Therefore, we can see the linear behaviour as some ``local" approximation of a potential with a more complicated shape, which would be bounded from below and well-behaved, contrary to a naive extrapolation of the linear scaling.

The same conclusion derived here for the negative slope can be obtained for a case with positive slope, positive $\phi_{\mathrm{ini}}$, but negative coupling to photons $c_{\phi\gamma}=-1$.

\subsection{Quadratic potential}\label{sec: Quadratic potential}
\begin{figure}
    \centering
\includegraphics[width=0.49\textwidth]{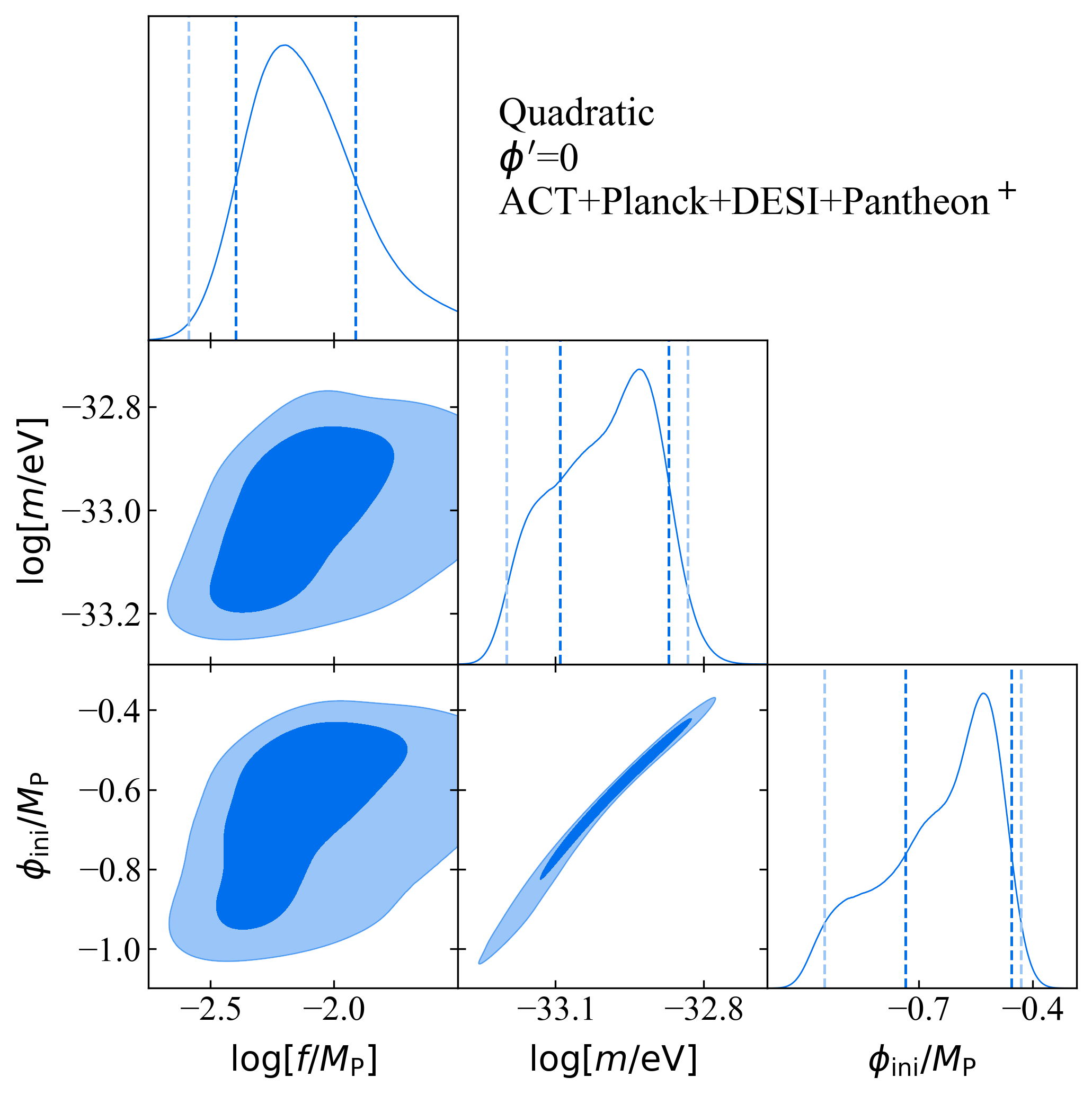}
\includegraphics[width=0.49\textwidth]{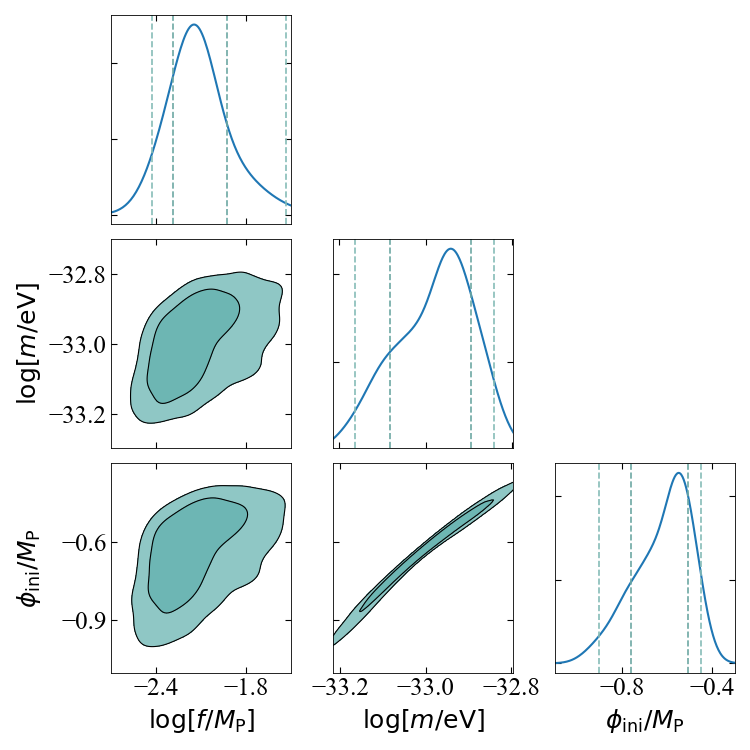}
    \centering
    \caption{Same as Fig.~\ref{fig:AL}, but for the case of the quadratic potential.}
    \label{fig:QU}
\end{figure}

We now consider a simple potential described by the quadratic form (again here $\mu=m)$:
\begin{equation}
    V(\phi)=\frac{1}{2}m^{2}\phi^2\;,
\end{equation}
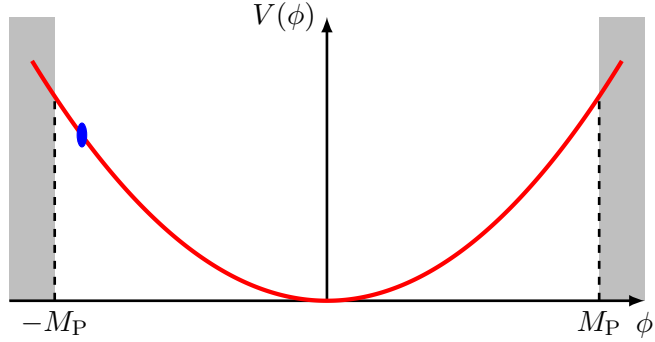
\begin{figure}[H]
    \centering
    \begin{tikzpicture}[xscale=0.6, yscale=1.5, >=latex]
    \fill[gray!50] (6, 0) rectangle (7, 2.5);
    \fill[gray!50] (-7, 0) rectangle (-6, 2.5);
    \draw[->, line width=1pt] (-7,0) -- (7,0) node[below] {$\phi$};
    \draw[->, line width=1pt] (0,0) -- (0,2.5) node[left] {$V(\phi)$};
    \draw[color=red, ultra thick, smooth, samples=200] 
        plot[domain=-6.5:6.5] (\x, 0.05*\x*\x);
    \node[below] at (-6,0) {$-M_{\mathrm{P}}$};
    \node[below] at (6,0) {$M_{\mathrm{P}}$};
    \draw[dashed, line width=1pt] (-6,0) -- (-6,1.8);
    \draw[dashed, line width=1pt] (6,0) -- (6,1.8);
    \filldraw[blue] (-5.4, 1.46) circle (3pt);
    \end{tikzpicture}
    \centering
    \caption{Quadratic potential.}
    \label{fig:Quadratic_Potential}
\end{figure}
which is often seen as the limit of the potential in Eq.~\ref{eq:cosV} when $\phi/f\ll 1$ (i.e., down the hill). However, this is not the case here, since $|\phi_{\mathrm{ini}}|/f\gtrsim 50$ (see Fig.~\ref{fig:QU}), and $\Delta\phi/\phi_{\mathrm{ini}}\lesssim 0.15 $ (see Eq.~\ref{eq:angleCB}), and thus the quadratic potential describes a different physical picture than the axion-like case of Sec.~\ref{sec: Axion-like potential}. Concerning DE probes, as already said, we probe only small field excursion but the slope away from zero is quite different between the quadratic and axion-like cases, and this also significantly impacts CB.
As for the axion-like potential we found that, for a rolling-down scenario, the initial kick is not necessary, and for simplicity now $ \phi_{\mathrm{kick}}^\prime$ is not included in the fit.

From Fig.~\ref{fig:QU} we can notice that the range of mass is again dictated by the preference for DE evolution. Then there is a strong degeneracy between the ALP mass and the initial displacement, because their combination provides the DE density. 
The parameter $f$ does not enter the quadratic potential, and thus it only affects the determination of the CB angle, where it is somewhat degenerate with the initial displacement, see Eq.~\ref{eq:angleCB} (we remind that here we set $c_{\phi\gamma}=1$).
Compared to the axion-like potential, here one gets the DE evolution preferred by observations for smaller $\Delta\phi$, in the case of large initial displacements, because of the stronger steepness of the potential. Therefore, the $f$ needed to explain CB is smaller, well below the Planck mass.
Flipping sign for $c_{\phi\gamma}$ and $\phi_{\mathrm{ini}}$ leads to an equivalent description, and, for the sake of conciseness, we do not include a ``freezing and thawing" case like Sec.~\ref{sec:kicked} because it does not improve the agreement with data with respect to the current picture.

\subsection{Ratra-Peebles potential}\label{sec: Ratra-Peebles potential}
In this section, we consider the Ratra-Peebles potential~\cite{Peebles:1987ek,Sola:2016hnq}. It is a potential that can alleviate the coincidence problem and can be interpreted as the power-law tail of an effective potential that incorporates inflation. Taking the power-law index equal to one, its form is given by: 
\begin{equation}
    V(\phi)=\mu^{4}\frac{f}{\phi}\;.
    \label{eq:VRB}
\end{equation}
\begin{figure}[H]
    \centering
    \begin{tikzpicture}[xscale=0.8, yscale=1.5, >=latex]
    \fill[gray!50] (6, 0) rectangle (7, 2.5);
    \draw[->, line width=1pt] (0,0) -- (7,0) node[below] {$\phi$};
    \draw[->, line width=1pt] (0,0) -- (0,2.5) node[left] {$V(\phi)$};
    \draw[color=red, ultra thick, smooth, samples=200] 
        plot[domain=0.1:7] (\x, 0.24/\x);
    \node[below] at (6,0) {$M_{\mathrm{P}}$};
    \draw[dashed, line width=1pt] (6,0) -- (6,1.8);
    \filldraw[blue] (0.2, 1.46) circle (3pt);
    \end{tikzpicture}
    \centering
    \caption{Ratra-Peebles potential.}
    \label{fig:RP_Potential}
\end{figure}
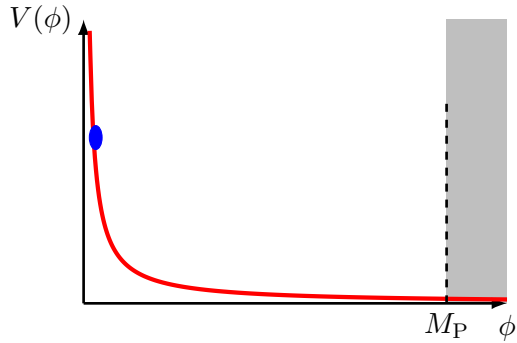
\begin{figure}
    \centering
\includegraphics[width=0.49\textwidth]{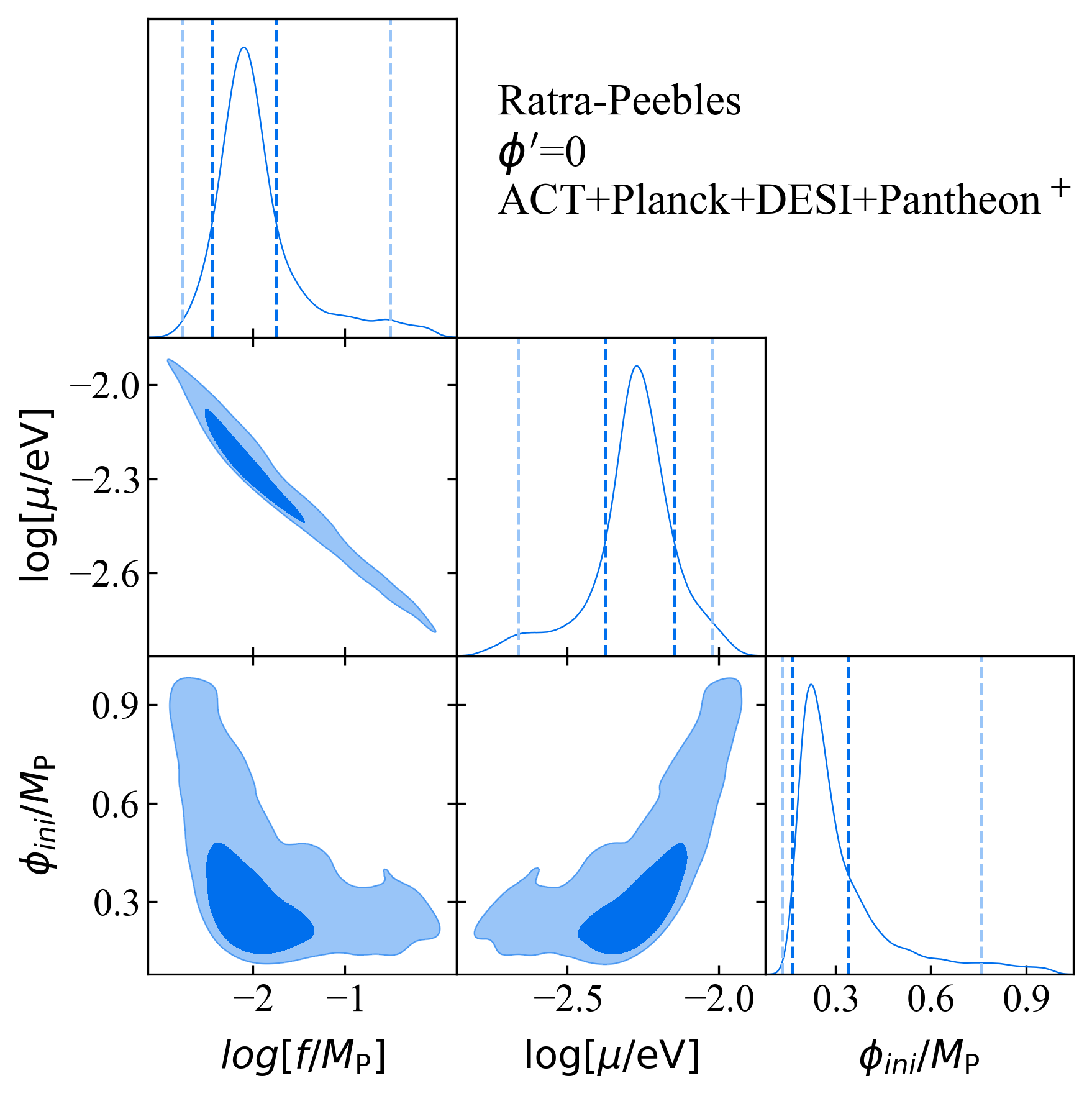}
\includegraphics[width=0.49\textwidth]{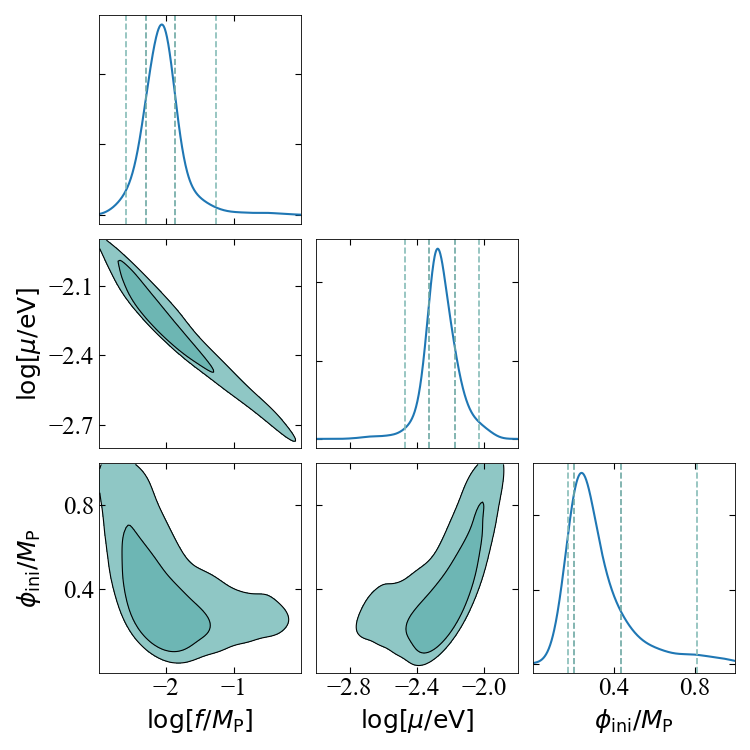}
    \centering
    \caption{Same as Fig.~\ref{fig:AL}, but for the case of the Ratra-Peebles potential.}
    \label{fig:RB}
\end{figure}

The physical picture is not very different from the case of the linear potential with negative slope, just the rolling is occurring with a different slope. 
The requirement of having $\mu^4\,f/\phi_{\mathrm{ini}}$ matching the DE density sets the degeneracy between $\mu^4\,f$ and $\phi_{\mathrm{ini}}$ in Fig.~\ref{fig:RB}.
By solving the equation of motion (which is now more complicated than in the linear case), one can see that a negative correlation between $\Delta\phi$ and $\mu^4\,f$ arises, and so also between $\Delta\phi$ and $\phi_{\mathrm{ini}}$.
Then a similar reasoning as before follows, i.e., large $\phi_{\mathrm{ini}}$ means small $\Delta\phi$, and thus small $f$ to fit CB.
Now, however, differently from the linear potential case, $f$ is in the numerator of Eq.~\ref{eq:VRB} and thus its degeneracy with $\mu$ goes in the opposite direction with respect to Fig.~\ref{fig:LIm}. The same applies also for the degeneracy between $\mu$ and $\phi_{\mathrm{ini}}$.

For the sake of conciseness, as in the quadratic case, we do not include a ``freezing and thawing” scenario, which increases the number of parameters without improving the likelihood value.

\subsection{Model comparison}\label{sec:comparison}
\begin{figure}[H]
    \centering
\includegraphics[width=0.47\textwidth]{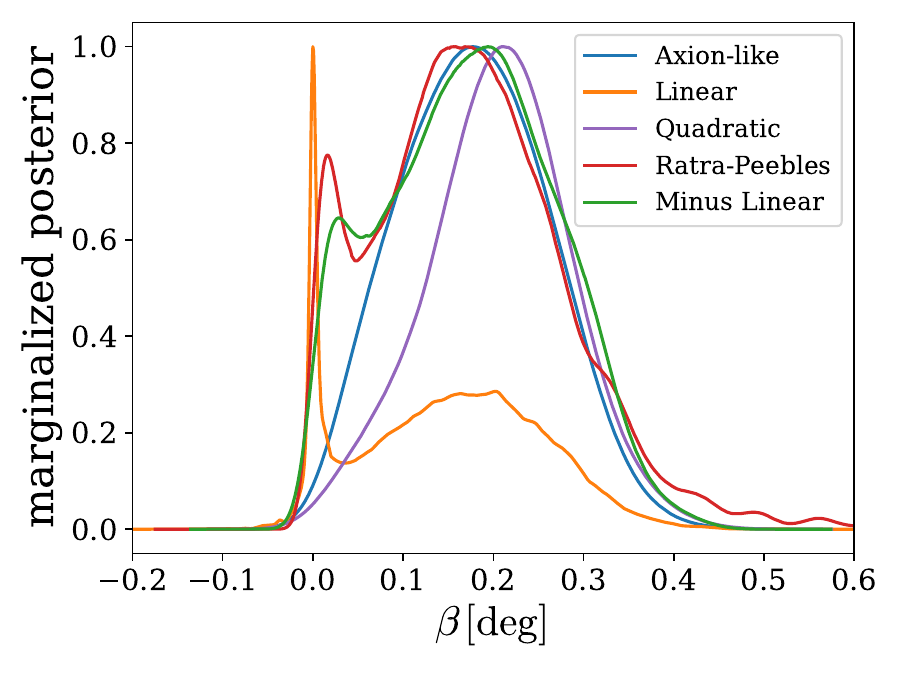}
\includegraphics[width=0.47\textwidth]{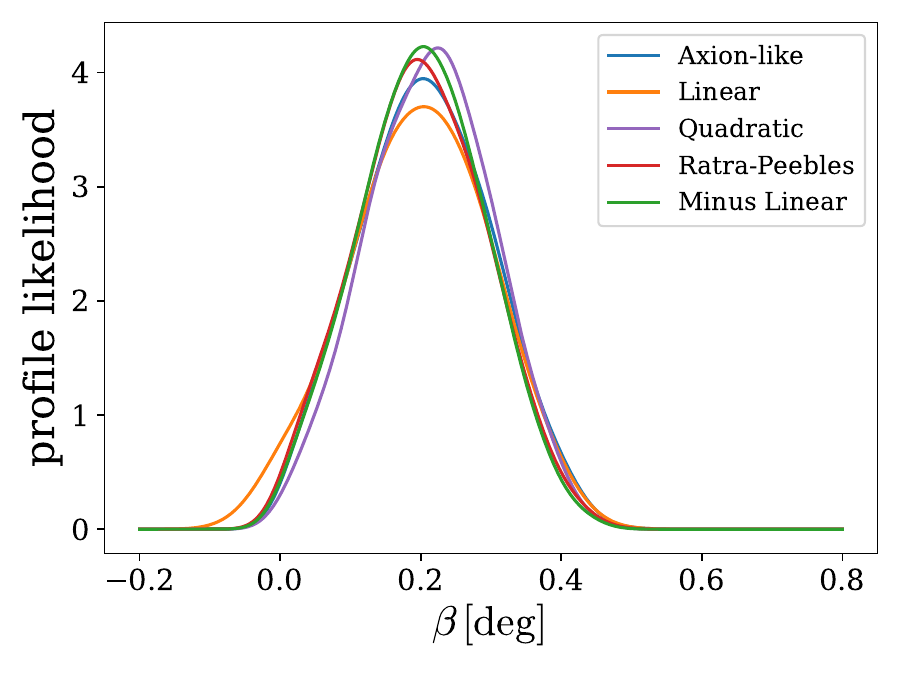}
    \centering
    \caption{Marginalized posterior (left) and profile likelihood (right) distributions of the cosmic birefringence $\beta$. The axion-like curves refer to the pure-thawing scenario.}
    \label{fig:betacomp}
\end{figure}

\begin{figure}[H]
    \centering
\includegraphics[width=0.47\textwidth]{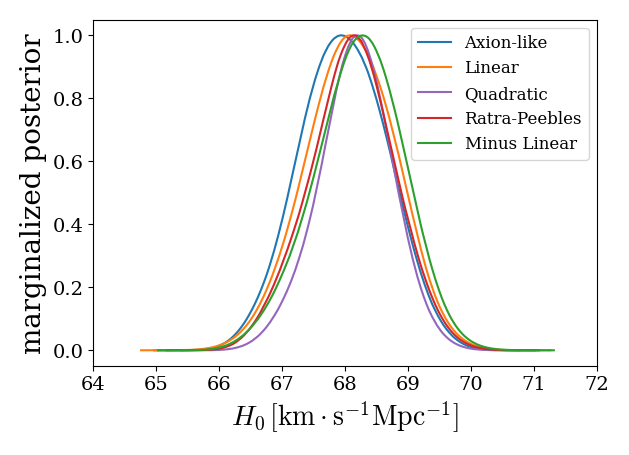}
\includegraphics[width=0.47\textwidth]{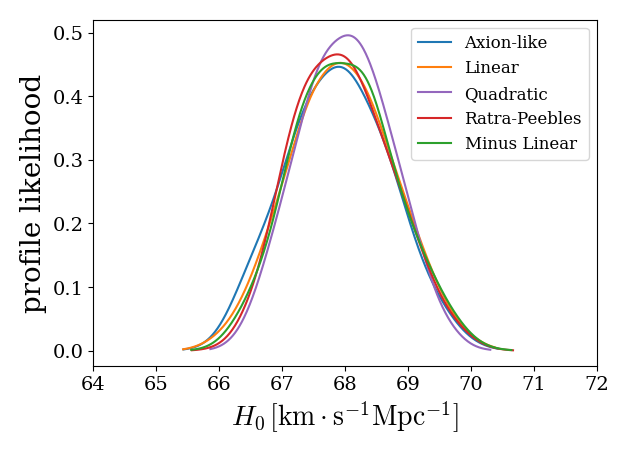}
    \centering
    \caption{Marginalized posterior (left) and profile likelihood (right) distributions of the Hubble rate today $H_0$. The axion-like curves refer to the pure-thawing scenario.}
    \label{fig:H0comp}
\end{figure}

From the distributions of the sampled parameters we can build the distribution of some derived parameters of interest. In Figs.~\ref{fig:betacomp} and \ref{fig:H0comp}, we show the posterior (left) and profile likelihood (right) distributions for the CB angle $\beta$ and for the Hubble rate today $H_0$. All the DE potentials considered essentially agree with each other and with expectations, for what concerns the best-fit scenario of both parameters. The mean (best-fit) values and 68\% credible (confidence) intervals are again reported in Table~\ref{table:params}. The Hubble parameter is found to be around 68 km/s/Mpc. 

In some scenarios, the marginalized posterior distribution of the CB angle in the left panel of Fig.~\ref{fig:betacomp} shows a peak at $\beta=0$. This is clearly not fitting the CB measurements, but since CB is just one data-point in the statistical analysis, thus essentially set by one parameter, there is a big prior volume effect which reduces its importance. This is made clear by the profile likelihood distribution in the right panel, which peaks at the measured value for all potentials, without showing a second peak at $\beta=0$.

\begin{figure}[H]
    \centering
\includegraphics[width=0.47\textwidth]{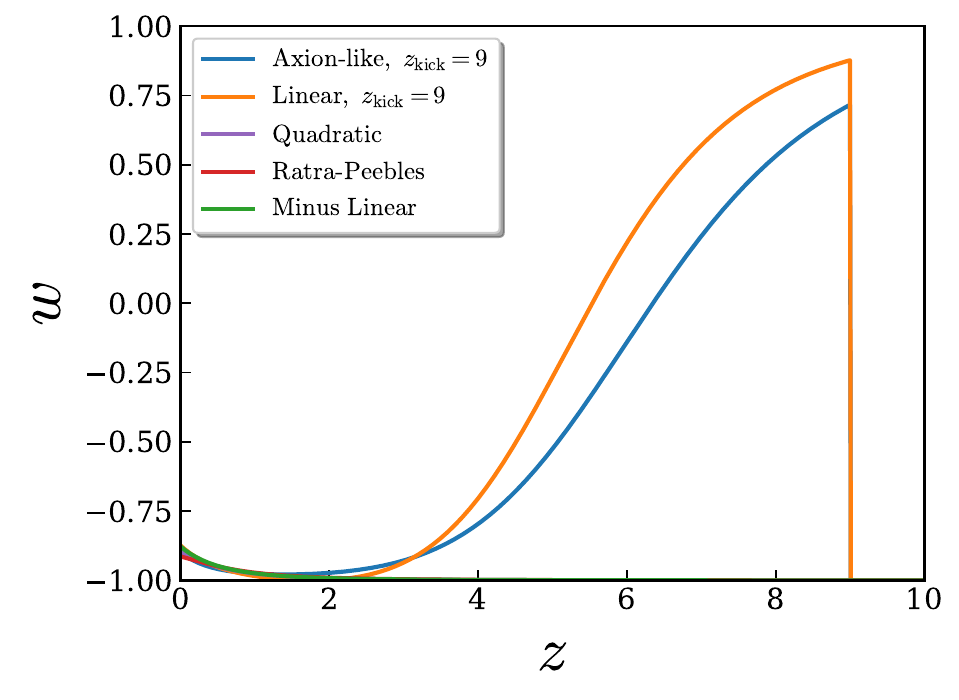}
\includegraphics[width=0.47\textwidth]{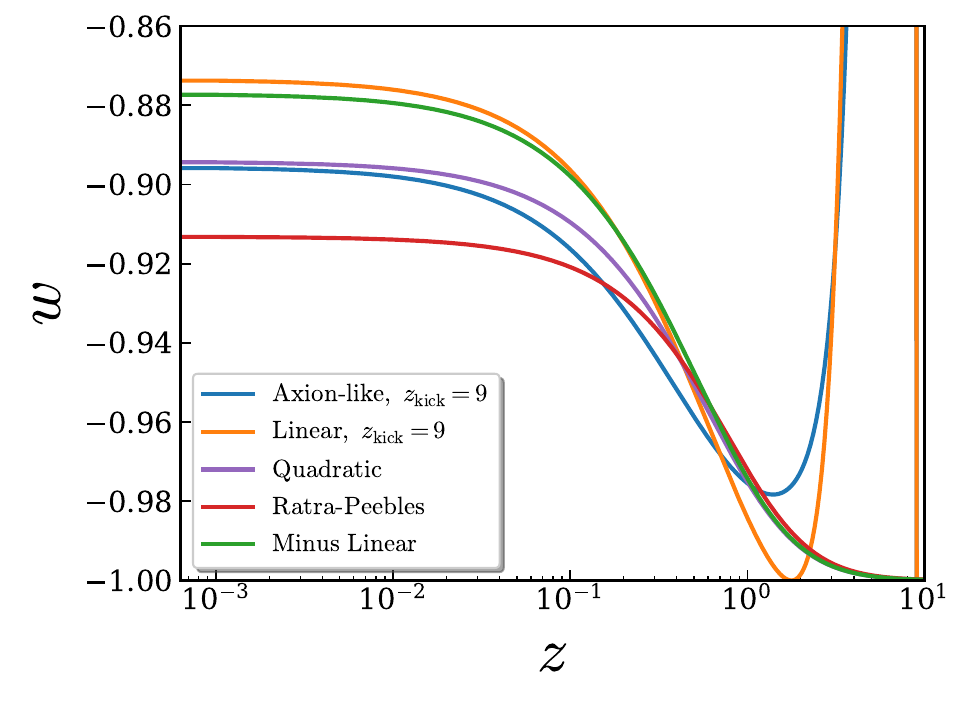}
    \centering
    \caption{Equation of state of the different potentials for pseudo-scalar DE. Left and right panels show the same figure but with different axes scales, to better illustrate the different behaviors.}
    \label{fig:w(z)}
\end{figure}

The evolution of the equation of state at $z\lesssim 2$ is similar in the various models, as shown in Fig.~\ref{fig:w(z)}, which reports the best-fit models of Table~\ref{table:params}. 
For the cases with no velocity kick, the field is frozen at $\omega=-1$ from the Hubble dragging down to very low redshift, then it starts its evolution.
In the cases with non-zero initial velocity, there is instead a significant evolution right after the kick, which can be relevant for CB but is negligible for the background dynamics of the Universe, since occurring at a time when DE is subdominant. To explain cosmological observables, it is crucial that DE reaches a slow-roll phase at late times, where then a significant kick velocity would not help.
From Fig.~\ref{fig:w(z)} we see that all models converge to $\omega\simeq -1$ at $z\approx2$, where a rolling towards $\omega\simeq -0.9$ starts, which is the behaviour required to fit DE cosmological data.

\begin{figure}[H]
    \centering
\includegraphics[width=0.47\textwidth]{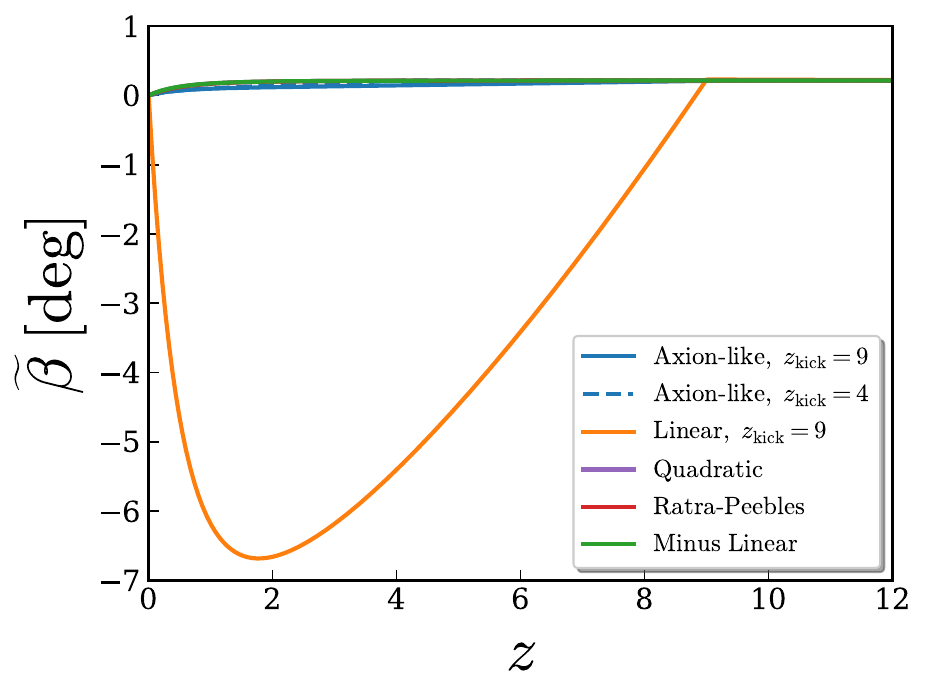}
\includegraphics[width=0.47\textwidth]{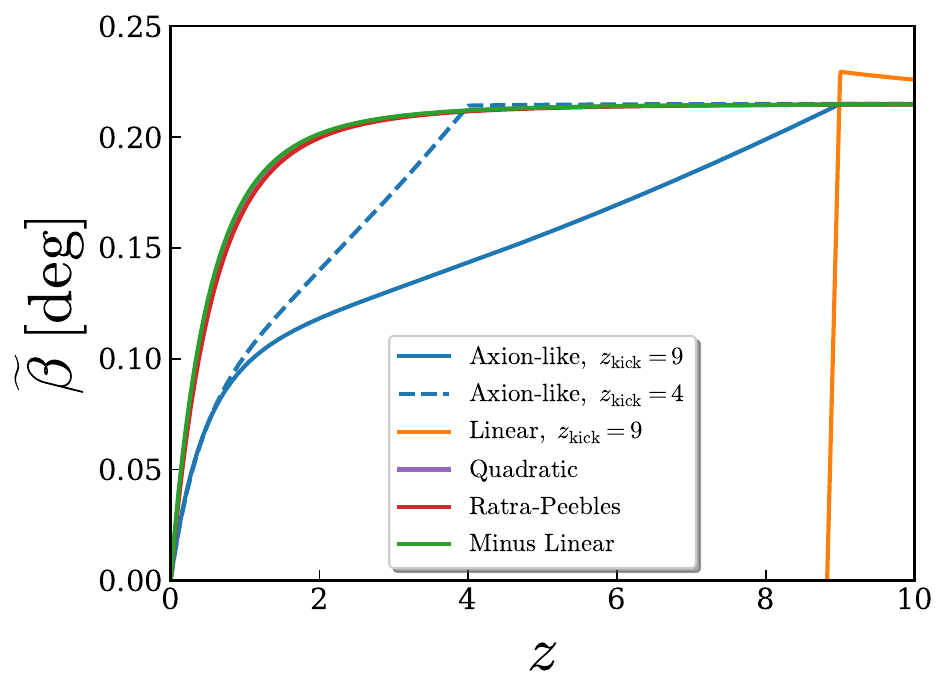}
    \centering
    \caption{Birefringence angle today as a function of the redshift of the photon source for different potentials of pseudo-scalar DE. To highlight the redshift evolution, we normalize the CB angle through $\tilde\beta(z)=\beta(z)\,\beta_{\rm data}/\beta_{\rm bf}$, where $\beta_{\rm data}=0.215$ is the measured value for the CMB~\cite{Diego-Palazuelos:2025dmh}, and $\beta_{\rm bf}$ is the best-fit CB angle reported in Table~\ref{table:params} for each model. Left and right panels show the same figure but with different y-axis scales, to better illustrate the different behaviours.}
    \label{fig:BETA(z)}
\end{figure}
In Fig.~\ref{fig:BETA(z)}, we report the dependence of the CB angle on the redshift for the best-fit of the different models. In the plot, an additional case for the axion-like potential in which the kick velocity is inserted at a smaller redshift ($z=4$, still before the beginning of the DE-dominated era) is included to illustrate how the CB evolution depends on the redshift of the kick. 
The reference choice for the redshift of the kick $z_{\rm kick} = 9$ was somewhat arbitrary, but since, after the kick, the equation of state deviates significantly from $\omega=-1$, this has to occur at $z\gg2$, in order not to violate DESI constraints. Qualitatively, different $z_{\rm kick} \gg 2$ do not produce significantly different cosmological scenarios. As discussed above, the field eventually relaxes back toward the cosmological constant limit and then increases again at very low $z$. The impact of the redshift of the kick on CB is instead more relevant, as can be seen in Fig.~\ref{fig:BETA(z)}.
The curves in the figure should be interpreted as the CB angle that an observer would measure today for polarized photons emitted from a source at redshift $z$ (i.e., it is not the birefringence angle that an observer would measure at redshift $z$). The right panel shows that the CB angle has nearly the same redshift behaviour in all the cases without a kick, while the axion-like and linear potentials where $\phi_{\mathrm{kick}}^\prime\neq 0$ could be easier to distinguish (if we are able to measure CB from sources at different redshifts). The CB evolution due to the velocity kick matches the behaviour of the equations of state plotted in Fig.~\ref{fig:w(z)}. The fact that in the case of the linear potential with positive slope the field first rolls up the potential, stops, and then rolls back down, translates into an inversion of the birefringence angle for a source located between us the redshift of the kick. In this case, the evolution of $\beta$ is rather strong, reaching values $|\beta|\gg1$ deg after the kick and for the majority of the Universe evolution. Thus, for this potential, having the small value of $\beta$ measured today from CMB photons looks somewhat a fine-tuned coincidence.

We now compare the results of the ALP DE scenarios discussed here with the case of the concordance model.
The latter includes a cosmological constant $\Lambda$ as DE and no pseudo-scalar DE. The improvement in the fit given by ALP DE with respect to $\Lambda$ is remarkable, with $\Delta\chi^2\simeq 14$ for what concerns the dataset without CB. In the last line of Table~\ref{table:params}, we compute the Akaike information criterion (AIC) and report $\Delta {\rm AIK_{c}}=\chi^2_{\rm cosmo,\Lambda}-\chi^2_{\rm cosmo,\phi DE}-2k_d$, where $k_d$ is the difference in the number of parameters between the pseudo-scalar DE model and $\Lambda$CDM. Evaluating $\exp(-\Delta {\rm AIK_{c}}/2)$ as the probability of $\Lambda$CDM compared to a model with pseudo-scalar DE, we find from 2.5\% for the axion-like case to 0.3\% for quadratic and Ratra-Peebles potential, with the difference among different models mainly driven by the number of parameters. We thus confirm the $\sim 3\sigma$ preference for dynamical DE found in other works in the literature.
Clearly, CB is not predicted by $\Lambda$CDM, and thus its inclusion adds another $\chi^2_{\beta,\Lambda}\simeq 8.4$ to the total $\Delta\chi^2$ (see Eq.~\ref{eq:likebeta}), bringing the evidence for pseudo-scalar DE close to $4\sigma$, e.g., we found $\exp(-\Delta {\rm AIK_{tot}}/2)\simeq 10^{-4}$ for the Ratra-Peebles case.

\begingroup
\renewcommand{\arraystretch}{1.5} 
\begin{table}
\small
\centering
  \begin{tabular}{|c|c|c|c|c|c|} 
\hline
\textbf{} & \textbf{Axion-like} & \textbf{Linear} & \textbf{Minus Linear} & \textbf{Quadratic} & \textbf{Ratra-Peebles} \\ 
\hline
    $\log [f/M_\mathrm{{P}}]$ & $-0.75^{+0.23}_{-0.11}$ & $-2.19^{+0.54}_{-0.83}$ & $-2.02^{+0.19}_{-0.46}$ & $-2.06^{+0.15}_{-0.34}$ & $-1.93^{+0.18}_{-0.50}$ \\
  & $-0.88^{+0.29}_{-0.13}$  & $-2.99^{+0.74}_{-0.51}$ & $-2.22^{+0.30}_{-0.12}$ & $-2.13^{+0.20}_{-0.15}$  & $-2.12^{+0.25}_{-0.17}$  \\
\hline
$\log [\mu/\mathrm{eV}]$ & / & $-3.06^{+0.14}_{-0.21}$ & $-3.03^{+0.10}_{-0.12}$ & / & $-2.28^{+0.14}_{-0.09}$ \\
 &  & $-3.24^{+0.18}_{-0.13}$ &$-3.05^{+0.08}_{-0.07}$  &  & $-2.27^{+0.09}_{-0.07}$  \\
\hline
$\log [m/\mathrm{eV}]$ & $-32.58^{+0.13}_{-0.22}$ & / &/  & $-33.00^{+0.13}_{-0.09}$ & /\\
 & $-32.59^{+0.34}_{-0.15}$ &  &  & $-32.95^{+0.06}_{-0.13}$  &  \\
\hline
$\phi'_{\mathrm{kick}}$ & $0.102^{+0.026}_{-0.102}$ & $0.177^{+0.073}_{-0.073}$ &/ & / & \\
$[{M_\mathrm{P}H_0}]$ & $0.113^{+0.061}_{-0.085}$  &  $0.154^{+0.085}_{-0.027}$ &  &  & \\
\hline
$\phi_{\mathrm{ini}}$ & $-0.43^{+0.13}_{-0.09}$ & $0.30^{+0.03}_{-0.16}$ & $-0.38^{+0.16}_{-0.02}$ & $-0.65^{+0.19}_{-0.09}$ & $0.32^{+0.02}_{-0.15}$ \\
 $[M_\mathrm{{P}}]$& $-0.35^{+0.07}_{-0.17}$ & $0.23^{+0.26}_{-0.04}$ & $-0.28^{+0.02}_{-0.21}$   &  $-0.56^{+0.05}_{-0.20}$ & $0.22^{+0.22}_{-0.01}$ \\
\hline
$\log c_{\phi\gamma}$ & $1.06^{+0.37}_{-0.28}$ & / & / & / &/ \\
 & $0.88^{+0.63}_{-0.04}$ &  &  &  & \\
\hline\hline
    $\beta$ & $0.18^{+0.09}_{-0.09}$& $0.15^{+0.11}_{-0.13}$ &$0.17^{+0.10}_{-0.10}$  &$0.18^{+0.09}_{-0.09}$  &$0.16^{+0.09}_{-0.16}$ \\ 
    $[\rm{deg}]$ &$0.16^{+0.12}_{-0.02}$  &$0.24^{+0.04}_{-0.11}$  & $0.26^{+0.01}_{-0.12}$  & $0.20^{+0.08}_{-0.06}$   &$0.23^{+0.04}_{-0.09}$\\
    \hline
    $H_0$ & $67.93^{+0.73}_{-0.71}$ & $68.09^{+0.73}_{-0.72}$ & $68.17^{+0.70}_{-0.70}$  &$68.14^{+0.70}_{-0.61}$  &$68.09^{+0.72}_{-0.66}$ \\
    $[\rm{km/s/Mpc}]$&$67.90^{+0.59}_{-0.63}$&$68.05^{+0.47}_{-0.72}$ &  $67.60^{+0.86}_{-0.39}$& $67.86^{+0.70}_{-0.38}$   & $67.65^{+0.80}_{-0.37}$\\
    \hline
    $\mathcal{\chi}^2_{\rm total}$ & $3848.7$& $3848.3$ &$3848.0$  &$3848.3$  &$3847.5$ \\
    \hline
    $\Delta {\rm AIK_{c}}$ & $6.9$& $8.7$ &$11.1$  &$10.7$  &$11.5$ \\
  \hline
\end{tabular}
  \caption{First five lines: Mean and 68\% credible interval (upper), plus best-fit and 68\% confidence interval (lower) for the parameters of the models sampled in the MCMC. Sixth and seventh lines: Same as for the previous lines, but for the derived parameter $\beta$ and $H_0$. 2nd last line: Value of the $\chi^2$ of the best-fit model for the different potentials, including both the cosmological and CB likelihoods. Using same data and MCMC procedure, the $\Lambda$CDM model gives $\chi^2_{\rm cosmo}=3861$ and $\chi^2_\beta=8.4$.
  Last line: $\Delta {\rm AIK_{c}}=\chi^2_{\rm cosmo,\Lambda}-\chi^2_{\rm cosmo,\phi DE}-2k_d$, where $k_d$ is the difference in the number of parameters between the pseudo-scalar DE model and $\Lambda$CDM, and only the cosmological likelihood is considered in the $\chi^2$ computation. The axion-like results refer to the pure-thawing scenario only.
   }
\label{table:params}
\end{table}
\endgroup

\section{Conclusions}\label{Sec:Conclusion}
In this work, we consider a pseudo-scalar field playing the role of dynamical DE. We analyze four different forms for its potential and different initial conditions for its evolution.
The reason to consider a pseudo-scalar field instead of a pure scalar field is (on top of avoiding fifth force constraints) to account for CB, i.e., for the rotation of the polarization plane of CMB photons, possibly observed in recent data~\cite{Diego-Palazuelos:2025dmh}. 

By comparing model predictions with observations from the CMB, SNIa and BAO, we found the evidence for pseudo-scalar DE over a cosmological constant is at the level of $\sim 3\sigma$, when considering only probes of the background expansion history, growing to $\sim 4\sigma$ when the CB is added.

In particular, we find that a (nearly) hilltop axion-like potential is a viable model but able to explain the cosmological evolution and CB at the same time only for large values of the anomaly coefficient, with $\log c_{\phi\gamma}\simeq 1$. If the hilltop condition is reached through a velocity kick rather than through the initial displacement, the required anomaly coefficient turns out to be even higher. 
The other potentials considered can work instead with $c_{\phi\gamma}= 1$. Linear models with positive slope and positive anomaly coefficient provide, however, a fine-tuned explanation of CB. Finally, scenarios in which the pseudo-scalar field rolls down a potential with quadratic, linear, or Ratra–Peebles forms look fully viable and with a symmetry-breaking scale close to the GUT scale.

We conducted the statistical analysis employing both a Bayesian and a frequentist approach, with the two methods providing very similar results.

Forthcoming cosmological observations, such as new data releases from DESI and Euclid~\cite{2011arXiv1110.3193L}, can significantly improve the assessment of DE evolution. 
At the same time, CMB experiments with absolute calibration of the polarization~\cite{BICEPKeck:2024cmk,Murata:2023heo,Coppi:2025fmt} will provide robust constraints on CB.
Combined together they might decide the fate of pseudo-scalar DE.

%%%%%%%%%%%%%%%%%%%%%%%%%%%%%%%%%
\section*{Acknowledgements}
%%%%%%%%%%%%%%%%%%%%%%%%%%%%%%%%%
We acknowledge N. Fornengo, M. Taoso, and T. Zanzarella for useful discussions throughout the development of the project.

The work of AM, MR and YW is supported by the European Union – Next Generation EU and by the Italian Ministry of University and Research (MUR) via the PRIN 2022 project n. 20228WHTYC - CUP D53C24003550006.
AM, MR and YW acknowledge support from the Research grant TAsP (Theoretical
Astroparticle Physics) funded by \textsc{infn}.

\bibliographystyle{bibi}
\bibliography{bibliography}

\end{document}